\pdfoutput=1

\documentclass[11pt,twoside,a4paper,cmspaper,final,collab]{cms-tdr}
\def\svnVersion{220877}\def\cmsCernNo{2013-153}\def\cmsCernDate{\today}\def\cmsMessage{Published in the Journal of High Energy Physics as \href{http://dx.doi.org/10.1007/JHEP12(2013)039}{\doi{10.1007/JHEP12(2013)039.}}}

\begin{document}\cmsNoteHeader{EWK-11-015}

\hyphenation{had-ron-i-za-tion}
\hyphenation{cal-or-i-me-ter}
\hyphenation{de-vices}
\RCS$Revision: 216633 $
\RCS$HeadURL: svn+ssh://svn.cern.ch/reps/tdr2/papers/EWK-11-015/trunk/EWK-11-015.tex $
\RCS$Id: EWK-11-015.tex 216633 2013-11-15 14:38:57Z chiochia $
\cmsNoteHeader{EWK-11-015} 

\title{Measurement of the cross section and angular correlations for
  associated production of a \cPZ~boson with b hadrons in pp collisions at $\sqrt{s}=7\TeV$}

\date{\today}

\abstract{
A study of proton-proton collisions in which two b hadrons are produced in
association with a \cPZ~boson is reported. The collisions were recorded at a
centre-of-mass energy of 7\TeV with the CMS detector at the LHC, for an integrated luminosity of $5.2\fbinv$. The b hadrons are identified by means of displaced secondary vertices,
without the use of reconstructed jets, permitting the study of
b-hadron pair production at small angular separation.
Differential cross sections are presented as a function of the angular
separation of the b hadrons and  the \cPZ~boson. In addition, inclusive measurements are presented. For both the inclusive and differential studies, different ranges of \cPZ~boson momentum are considered, and each measurement is compared to the predictions from different event generators at leading-order and next-to-leading-order accuracy.
}

\hypersetup{%
pdfauthor={CMS Collaboration},%
pdftitle={Measurement of the cross section and angular correlations for associated production of a Z boson with b hadrons in pp collisions at sqrt(s) = 7 TeV},%
pdfsubject={CMS},%
pdfkeywords={CMS, physics, b hadrons, angular correlations}}

\maketitle 

\section{Introduction\label{sec:intro}}

The measurement of \cPZ$/\cPgg^*$ (henceforth denoted by ``\cPZ'') production in association with b quarks
at the Large Hadron Collider (LHC) is relevant
for various experimental searches. In particular, the process constitutes one of the dominant backgrounds to
standard model (SM) Higgs boson production associated with a \cPZ~boson, where the Higgs boson
decays subsequently to a $\bbbar$ pair. The discovery by the ATLAS and
Compact Muon Solenoid (CMS) experiments
of a neutral boson with a mass of about
$125\GeV$~\cite{Aad20121,:2012gu} motivates further studies to
establish its nature and determine the
coupling of the new boson to b quarks.
Furthermore, for models featuring an extended Higgs sector, such as
two-Higgs-doublet models~\cite{Gerard:2007kn,deVisscher:2009zb,Dermisek:2008id,Dermisek:2008uu}, an interesting discovery channel is $\phi_1\rightarrow\cPZ\phi_2$ with the subsequent decay
$\phi_2 \rightarrow \bbbar$, where $\phi_{1,2}$ are neutral Higgs
bosons. Since the mass difference $m_{\phi_1}-m_{\phi_2}$ may be large, the Higgs decay would consist of a pair of collinear b quarks produced in association with a \cPZ~boson.

Of particular interest is the measurement of angular correlations of b
hadrons, especially at small opening angles, where significant theoretical uncertainties in the
description of the col\-lin\-e\-ar production of b quarks  remain.  Several
theoretical predictions, obtained with different techniques and
approximations, can be tested.
Tree-level calculations allowing for large numbers of extra partons in the matrix elements (as initial- and final-state radiation)
are available. These are provided by
\MADGRAPH~\cite{Alwall:2007st,Alwall:2011uj}, {\ALPGEN~\cite{alpgen},
  and \SHERPA~\cite{Gleisberg:2008ta},
in both the five- and  four-flavour approaches, \ie by considering the
five or four lightest quark flavours in the proton parton distribution
function (PDF) sets. Next-to-leading-order
(NLO) calculations have been performed in both the five-flavour
(\MCFM)~\cite{Campbell:2000bg} and four-flavour~\cite{FebresCordero:2008ci,Cordero:2009kv}  approaches.
A fully automated NLO computation matched to a parton shower simulation is
implemented by the a\MCATNLO event
generator~\cite{Frederix:2011qg,Frixione:2002ik}.
A detailed discussion of b-quark production in the different calculation schemes is available in Ref.~\cite{Maltoni:2012pa}.

From the experimental point of view, the study of b-hadron pair
production using the standard jet-based b-tagging
methods~\cite{Chatrchyan:2012jua} suffers from geometrical
limitations due to the jet cone size.  Hadronic cascades from
b-quark pairs at small angular separation can merge into a single jet,
making this region of phase space difficult to access using jet-based
b-tagging techniques. To overcome this obstacle, an alternative method
is used, consisting of the identification of b hadrons from displaced
secondary vertices, which are reconstructed from their charged decay
products. This approach is implemented in the inclusive secondary
vertex  finder (IVF)~\cite{Khachatryan:2011wq}. The IVF exploits
the excellent tracking capabilities of the CMS detector and, being
independent of the jet reconstruction, extends the sensitivity to small angular separations and softer b-hadron transverse momenta ($\pt$).

Four variables are used to parametrise the angular correlations in the $\cPZ\bbbar$ final state: $\Delta R_{\mathrm{BB}}$, $\Delta\phi_{\mathrm{BB}}$, min$\Delta R_{\mathrm{ZB}}$, and $A_{\mathrm{ZBB}}$.
The angular correlation between the b hadrons is described by two variables, $\Delta R_{\mathrm{BB}}$ and
$\Delta\phi_{\mathrm{BB}}$, the angular separation between the flight
directions of the two particles in $(\eta,\phi)$ and in the transverse plane, respectively.
The variable $\Delta R_{\mathrm{BB}}$ is defined as $\Delta
R_{\mathrm{BB}} = \sqrt{ (\Delta \phi_{\mathrm{BB}})^2 + (\Delta
  \eta_{\mathrm{BB}})^2}$, where $\Delta \phi_{\mathrm{BB}}$ and
$\Delta \eta_{\mathrm{BB}}$ are the azimuthal (in radians) and pseudorapidity
separations. The pseudorapidity is defined as $\eta=
-\ln[\tan(\theta/2)]$, where $\theta$ is the polar angle  relative to
the anticlockwise beam direction. The $\Delta R_{\mathrm{BB}}$
distribution constitutes a direct test of the modelling of the
different $\mathrm{pp}\rightarrow \cPZ \bbbar\mathrm{X}$ production modes. This quantity allows the identification of the contribution from the
$\cPq{i} \rightarrow \cPZ \bbbar \mathrm{X}$ subprocesses (where
$i=\cPq$, $\cPg$) for which the scattering amplitude modelling is
based on Feynman diagrams with $\cPg \rightarrow \bbbar$ splitting.
Leading order diagrams for these subprocesses are shown in
Figs.~\ref{diag} (a) and (b), together with diagrams representative of other
$\mathrm{pp}\rightarrow\cPZ\bbbar$ production modes: emission of a \cPZ~boson from a b-quark
line (c), and b-quark fusion $\cPg\cPg \rightarrow \cPZ\bbbar$
(d).
\begin{figure}[!htbp]
	\begin{center}
	\includegraphics[width=0.8\textwidth]{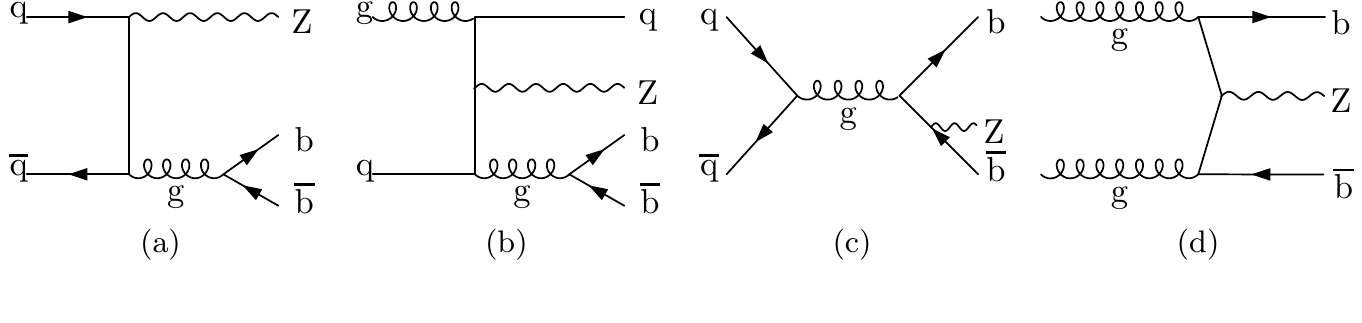}
	\caption{Tree-level Feynman diagrams for (a,b) $\cPq
          {i}\rightarrow \cPZ\bbbar \mathrm{X}$ subprocesses
          (where $i=\mathrm{q,g}$) involving $\cPg\rightarrow \bbbar$
          splitting; (c)  $\cPq\cPaq \rightarrow \cPZ\bbbar$ with the
          emission of a $ \cPZ$~boson from a b quark; and (d) $\cPg\cPg \rightarrow \cPZ\bbbar$.}
\label{diag}	
	\end{center}
\end{figure}

A second variable, the angular separation between the b hadrons in the transverse plane,
$\Delta\phi_{\mathrm{BB}}$, is also considered
because it is a better observable for the back-to-back configuration. Since the relative fraction
of quark- and gluon-initiated subprocesses is correlated with the \cPZ-boson momentum $\pt^{\cPZ}$, the differential $\Delta R_{\mathrm{BB}}$ and $\Delta\phi_{\mathrm{BB}}$ distributions are measured in different intervals of $\pt^{\cPZ}$.

Two additional angular variables are considered:
the angular separation between the \cPZ~boson and the closest b hadron in
the $(\eta,\phi)$ plane, min$\Delta R_{\mathrm{ZB}}$, and the asymmetry between the b-hadron emission directions and the \cPZ~production direction, $A_{\mathrm{ZBB}}$, defined as
\begin{equation}
A_{\mathrm{ZBB}}=\frac{\text{max}\Delta R_{\mathrm{ZB}}-\text{min}\Delta R_{\mathrm{ZB}}}{\text{max}\Delta
R_{\mathrm{ZB}}+\text{min}\Delta R_{\mathrm{ZB}}},
\label{eq:Asym}
\end{equation}
where max$\Delta R_{\mathrm{ZB}}$ is the distance between the \cPZ~boson and the further b hadron.
Configurations in which the two b hadrons are emitted symmetrically with
respect to the \cPZ~direction yield a value of $A_{\mathrm{ZBB}}$ close to zero.
Emission of additional gluon radiation in the final state results in a nonzero
value of $A_{\mathrm{ZBB}}$. Hence, the $A_{\mathrm{ZBB}}$ variable helps
to indirectly test the validity of quantum chromodynamics (QCD) at higher orders of the perturbative series.
The min$\Delta R_{\mathrm{ZB}}$ variable identifies events with the \cPZ~boson in the vicinity of one of the two b hadrons, and is therefore useful for testing NLO corrections involving \cPZ~radiation from a quark~\cite{Rubin:2010xp}.

The contribution of the $\cPq i\rightarrow \cPZ\bbbar \mathrm{X}$
subprocesses to the total production is
illustrated  in Fig.~\ref{plotth} as a function of each of the four variables described above. The
distributions are shown for both the nonboosted (all $\pt^\cPZ$) and the boosted
($\pt^\cPZ>50\GeV$) regions of the \cPZ~transverse momentum.
For all the variables, the contribution of  the
$\cPq i\rightarrow \cPZ\bbbar \mathrm{X}$ subprocesses differs
from the
contribution of $\cPg\cPg \rightarrow \cPZ\bbbar \mathrm{X}$. The $\cPq i \rightarrow
\cPZ\bbbar \mathrm{X}$ subprocesses are dominant in the following regions: $\Delta
R_{\mathrm{BB}}<1$, $\Delta\phi_{\mathrm{BB}}<0.75$, min$\Delta
R_\mathrm{ZB}>3.2$, and $A_\mathrm{ZBB}<0.05$.
\begin{figure}[!htbp]
	\begin{center}
          \includegraphics[width=0.65\textwidth]{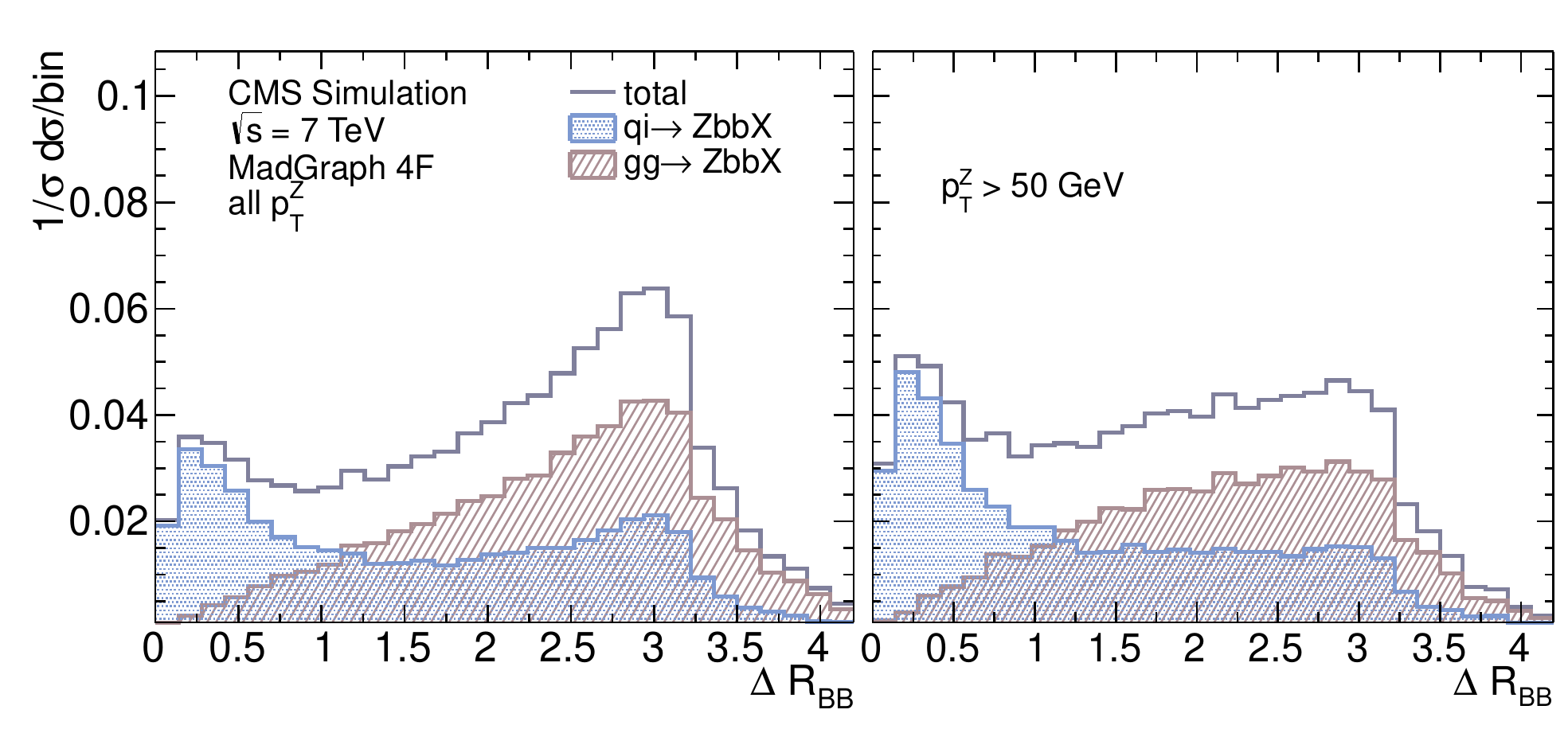}
          \includegraphics[width=0.65\textwidth]{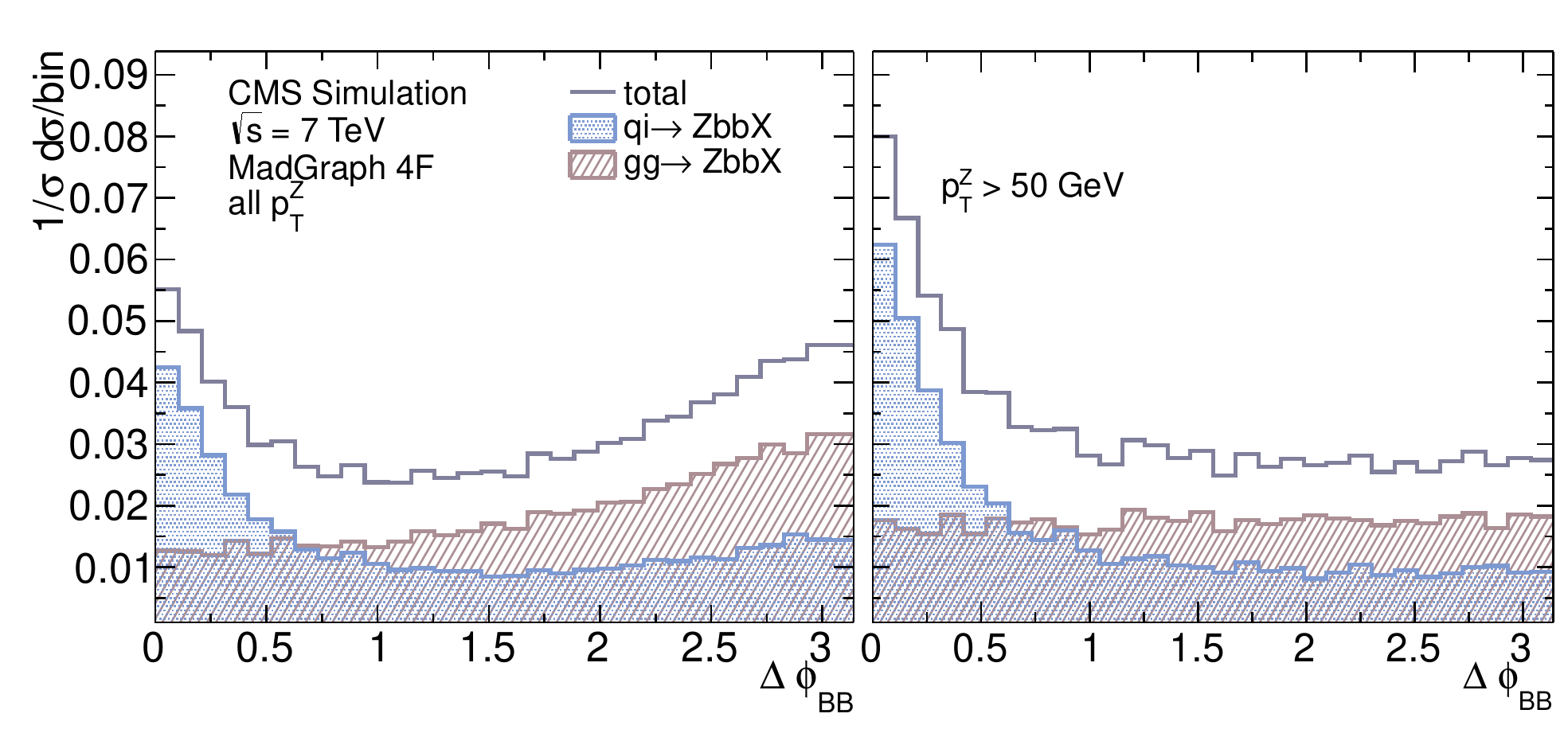}
	\includegraphics[width=0.65\textwidth]{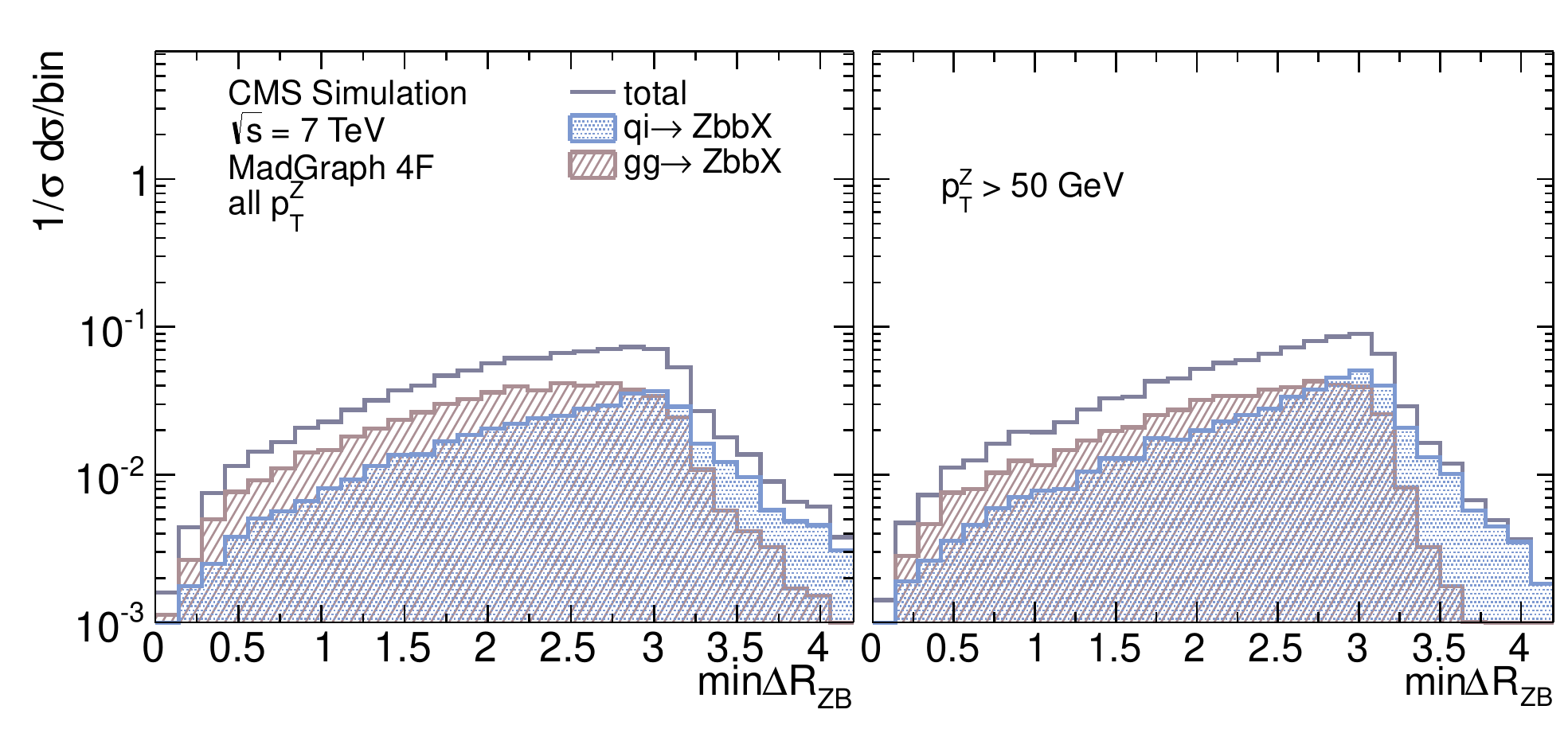}
        \includegraphics[width=0.65\textwidth]{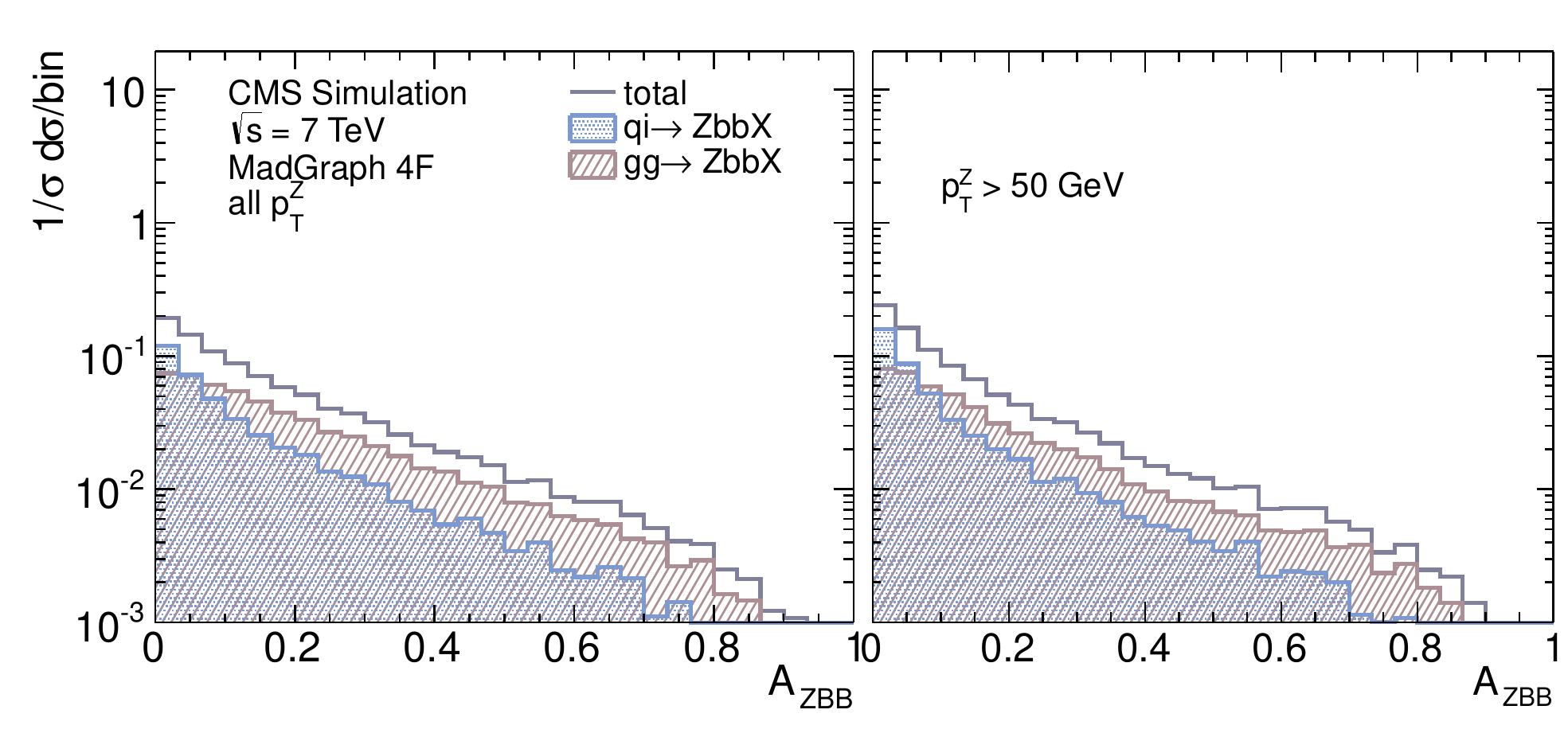}
	\caption{Distribution of $\Delta R_{\mathrm{BB}}$ (first row),
          $\Delta \phi_{\mathrm{BB}}$ (second row),   min$\Delta
          R_\mathrm{ZB}$ (third row), and $A_{\mathrm{ZBB}}$ (fourth
          row) as predicted by \MADGRAPH in the four-flavour
scheme, in the nonboosted (left) and boosted (right) regions of the \cPZ~
transverse momentum.
The component from $\cPg\cPg\rightarrow \cPZ\bbbar \mathrm{X}$ is
represented by the hatched histogram, while the contribution from
$\cPq i\rightarrow\cPZ\bbbar\mathrm{X}$ subprocesses
(where $i=\mathrm{q,g}$) is represented by
the shaded histogram. The unshaded histogram corresponds to the sum of the
two components. }
\label{plotth}	
	\end{center}
\end{figure}

In this analysis, the differential production cross sections for the
process $\mathrm{pp} \to
\cPZ\bbbar\mathrm{X}$ (henceforth the processes are denoted by their final state, here ``$Z\bbbar$'') as
functions of the four kinematic variables listed above are evaluated
from CMS data.  These cross sections are given at the hadron level and compared to the predictions provided by several of the Monte Carlo (MC) generators mentioned above.
The total cross section is also measured.
The results are given for different regions of $\pt^{\cPZ}$.
Because of the limited size of the available data sample, the differential measurements are
calculated in the nonboosted and boosted regions.
The total cross section is evaluated for $\pt^{\cPZ}$ larger than 0,
40, 80, and 120\GeV.
\cPZ~bosons are reconstructed in the $\Pep\Pem$ and $\mu^+\mu^-$ decay modes.
The analysis exploits the full 2011 data set recorded at $\sqrt{s}=7\TeV$, corresponding to an integrated luminosity of $(5.2\pm 0.1)\fbinv$.
Measurements of the \cPZ-boson production cross section in association with one
or two b-tagged jets at the LHC have been reported previously by the ATLAS and CMS Collaborations~\cite{Aad:2011jn,Chatrchyan:2012vr}.

The paper is organised as follows: the description of the CMS experiment and simulated
samples are given in
Section~\ref{sec:CMS_MC}; the event reconstruction and selection are presented in
Section~\ref{sec:eventsel}; the measurement technique is
explained in Section~\ref{sec:xsec}; the systematic uncertainties are
discussed in Section~\ref{sec:systematics}; the theoretical
uncertainties associated with different models of $\cPZ\bbbar$
production are summarized in Section~\ref{sec:therrors}; the results and
conclusions are presented in Sections~\ref{sec:results} and
\ref{sec:conclusions}, respectively.

\section{CMS detector and simulated samples
\label{sec:CMS_MC}}

A detailed description of the CMS experiment can be found in
Ref.~\cite{Chatrchyan:2008aa}. The main subdetectors used in this analysis are the
silicon tracker, the electromagnetic calorimeter (ECAL), and the muon system. The
tracker consists of silicon pixel and strip detector modules and is
immersed in a 3.8\unit{T} magnetic field, which enables the measurement of
charged particle momenta over the pseudorapidity range $\abs{\eta}<2.5$. The
electromagnetic calorimeter consists of nearly $76\,000$ lead
tungstate crystals, which provide coverage for
$\abs{\eta} \lesssim1.48$ in a cylindrical barrel region and $1.48
\lesssim \abs{\eta} \lesssim3.0$ in
two endcap regions, except for a insensitive gap in the region $1.442 < \abs{\eta} < 1.566$ between the
ECAL barrel and endcap.  Muons are
identified in the range  $\abs{\eta}<2.4$ by gas-ionisation detectors
embedded in the steel return yoke. The first level of the CMS trigger
system consists of custom hardware processors and uses information
from the calorimeters and muon system to select the most interesting
events in less than 1\mus. The high level trigger processor
farm further decreases the event rate to less than 300\unit{Hz} before data
storage.

Samples of signal and background events are produced using
various event generators to estimate the
signal purity, efficiency, and detector acceptance, with the CMS detector response modelled in
extensive detail with \GEANTfour~\cite{Agostinelli:2002hh}.

The  $\cPZ\bbbar$ signal sample is produced with the  \MADGRAPH 1.4.8
generator  in the four-flavour approach. No b quarks are present in the initial
state, while up to two additional light partons are produced in
association with the \cPZ~boson and the two b quarks. The PDF set is
CTEQ6L1 and the simulation of parton shower, hadronisation, and
multiparton interactions is done with \PYTHIA 6.4.2.4
~\cite{Sjostrand:2006za}. The background samples are \cPZ~plus jets, where
the additional  jets are from light quarks or gluons (u, d, c, s, g),
top pair production (\ttbar), and $\cPZ$~pair production.
The $\cPZ+\text{jets}$ sample is extracted from a Drell--Yan inclusive sample
produced with \MADGRAPH in the five-flavour approach and interfaced
with \PYTHIA. The \ttbar sample is also produced with the \MADGRAPH
generator interfaced with \PYTHIA, while the diboson
\cPZ\cPZ~sample is generated with \PYTHIA. The tune considered in
\PYTHIA is Z2$^*$, which is the Z1 tune~\cite{Field:2010bc} with the
PDF set changed to CTEQ6L1 and minor modifications of the underlying event modelling, namely $\texttt{PARP(90)}=0.227$ and $\texttt{PARP(82)}=1.921$.

Additional interactions per bunch crossing (pileup) are included
in the simulation with the distribution of pileup interactions
matching that observed in data.

\section{Event reconstruction and selection \label{sec:eventsel}}

The first step of the analysis is the online event selection with the loosest available dimuon and dielectron triggers in order to enrich the sample with $\cPZ\to \mu^+ \mu^-$ and  $\Pep\Pem$ decays.  The dielectron trigger line requires loose electron identification and isolation and imposes 17 and 8\GeV transverse momentum thresholds on the two electron candidates, respectively. The transverse momentum thresholds of the muon trigger line, which changed with time to cope with increasing instantaneous luminosity, were initially 7\GeV on both muon candidates, then 13 or 17\GeV on one candidate and 8\GeV on the other.

Muon candidates are then required to pass tight selection requirements to ensure high purity~\cite{Chatrchyan:2012xi}.
Electron candidates are reconstructed from energy deposits in the ECAL, and must satisfy the standard CMS electron identification criteria~\cite{Khachatryan:2010xn}. Leptons are required to have $\pt >20\GeV$, and to be within the
pseudorapidity range $\abs{\eta}<2.4$.
Prompt leptons are
selected by requiring a distance of closest approach
between the track and the primary pp interaction (identified as the vertex with the largest quadratic sum of its constituent tracks' \pt) smaller than 200\micron.
A requirement is applied on the lepton isolation, computed using the particle-flow
technique~\cite{CMS:2009nxa}, which exploits the information from all subdetectors to
individually identify the particles produced in the collisions.
The isolation, defined as the ratio between the scalar sum of the transverse momentum or transverse energy ($E_T$)
of the particles within a $\Delta R < 0.4~(0.3)$ cone around the
muon (electron) and its transverse momentum,
$ (\sum_{\text{charged had.}} \pt +
\sum_{\text{neutral had.}} E_T + \sum_{\text{photon}}
E_T)/\pt$,  must be at most 0.15.
In order to ensure that the selection is stable regarding the large and varying number of primary interactions, the charged
particle-flow candidates are required to be associated with the
selected primary vertex (PV).
In addition, a correction is applied to subtract the energy
contribution of neutral hadrons and
photons produced in pileup interactions. This correction is estimated
event by event from the median of the energy density distribution and applied within the isolation cone~\cite{Cacciari:2007fd}.

Only events with two oppositely charged same-flavour lepton candidates with invariant mass between  60 and 150\GeV are selected. The signal region is then defined as the  $81<{M}_{\ell \ell}<101\GeV$ interval to reduce the contamination from \ttbar events.

Events containing b hadrons are selected by applying the inclusive vertex finder technique.
The secondary vertex (SV) reconstruction on which the IVF is based is initiated by the identification of a set of
``seed'' tracks that are significantly displaced with respect to the primary vertex.
Such tracks are selected by requiring their three-dimensional impact
parameter to be larger than 50\micron, and their impact parameter
($\mathrm{IP}$) significance $S_\mathrm{IP}=\mathrm{IP}/\sigma_\mathrm{IP}$ larger than 1.2, where
$\sigma_{IP}$ is defined from the uncertainties on both the PV
position and the point of closest approach between the track and the PV.
Additional tracks are clustered together with the seed tracks if they
fulfil several requirements. First, the distance of closest approach
of a track to the seed must not exceed 500\micron, and its significance must be smaller than 4.5. Second, the angle between the vector defined by the PV and the point of closest approach on the seed track and the seed track direction at the vertex has to be smaller than 45$^{\circ}$ so only forward  tracks from b-hadron decays are retained. Secondary vertices are built from the seeds and clustered tracks~\cite{Fruhwirth:2007hz}.

The SV four-momentum is calculated as $p_\mathrm{SV}= \sum p_i$ where the sum
is over all tracks associated with that vertex.
The pion mass hypothesis is used for every track to obtain its energy $E_i$. The vertex mass $m_\mathrm{SV}$ is given by $m_\mathrm{SV}^2 = E^2_\mathrm{SV}- p^2_\mathrm{SV}$.

The IVF technique establishes a list of b-hadron (B) candidates from the reconstructed SVs.
If  two SVs are present, they can potentially be the signature of a $\cPqb \to
\cPqc\mathrm{X}$ decay chain and are merged into a single B
candidate if the following conditions are fulfilled:
i)~$\Delta R(\mathrm{SV}_1,\mathrm{SV}_2)<0.4$, ii)~the sum of the invariant masses of track candidates associated with the vertices
is smaller than 5.5\GeV, and iii)~$\cos\delta> 0.99$, where $\delta$ is the angle between the vector
from the position of the SV that is closer to the PV to the position
of the other SV and the three-momentum of the vertex with larger decay
length. The flight distance significance of a B candidate is
calculated from the distance between the PV and SV divided by its uncertainty.
More details of the SV and B candidate reconstruction can be found in
Ref.~\cite{Khachatryan:2011wq}.
\begin{figure}[!htbp]
	\begin{center}
\includegraphics[width=0.45\textwidth]{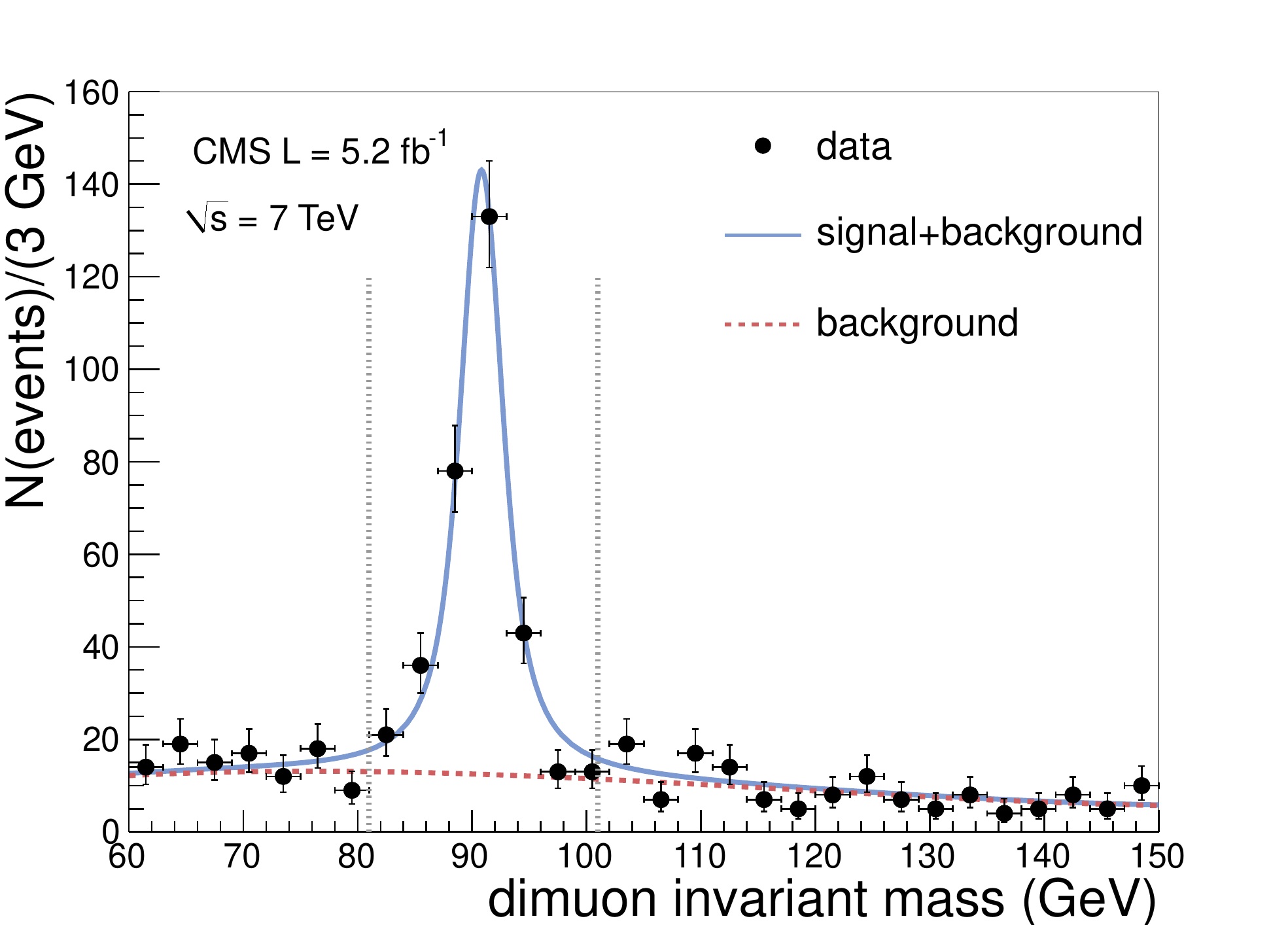}
\includegraphics[width=0.45\textwidth]{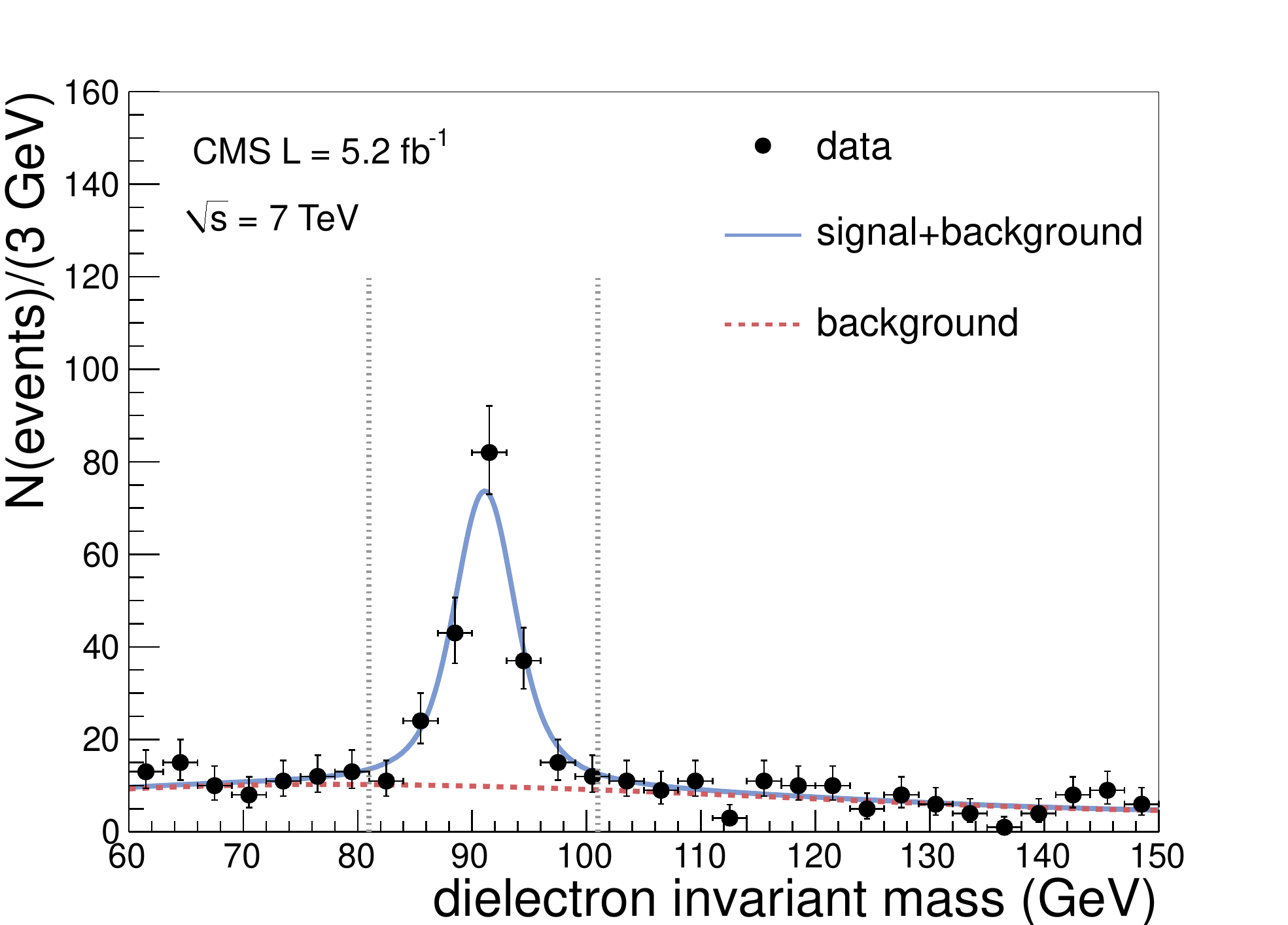}
	\caption{
Fit results for the dimuon (left) and dielectron (right) invariant
mass distributions for events with two leptons and two B candidates
selected as described in Section~\ref{sec:eventsel}.
The dashed line shows the fitted background component and the
solid line the sum of the fitted signal and background components,
which are described in the text. The vertical dashed lines indicate the
boundaries of the signal region. The points with errors represent the data.
\label{fig:dileptonmassfit}}	
	\end{center}
\end{figure}

The flight distance $L$ is defined as the length of the
three-dimensional vector connecting the primary and secondary
vertices. Its significance $S_L$ is obtained by dividing $L$ by its
uncertainty, calculated as quadratic sum of the PV and SV position
uncertainties.  A b hadron candidate is retained if $S_L>5$,
$\abs{\eta}<2$, $\pt>8\GeV$, and invariant mass $m>1.4\GeV$.
The B candidate mass and flight distance significance cuts, along with the requirement of at least three
tracks associated with the secondary vertex, are the most effective requirements
for rejecting background events from $\cPZ\ccbar$ production.

Events that have exactly two B candidates are retained. The resulting
dimuon and dielectron invariant masses are shown between 60 and
150\GeV in Fig.~\ref{fig:dileptonmassfit}. In total, 330 (223) events
pass all the selection requirements in the muon (electron) channel in
the $81<M_{\ell \ell}<101\GeV$ signal mass region.
Thanks to the excellent performance of the CMS tracking system, the
IVF angular resolution is approximately 0.02 for  $\Delta
R_{\mathrm{BB}}$ and $\Delta\phi_{\mathrm{BB}}$ and 0.03 for min$\Delta R_{\mathrm{ZB}}$ and $A_{\mathrm{ZBB}}$.

The main source of background contamination in the final sample is top-quark pair production. The
\ttbar fraction is assessed from an unbinned maximum-likelihood fit to the measured dilepton invariant mass distribution as
described in Section~\ref{sec:xsec}.
The fit yields a \ttbar contamination of approximately 30\% in the inclusive event sample, and of about 23\% for $\pt^\cPZ >50\GeV$.

The measured and simulated distributions of the most significant
event properties are compared at the detector level, as shown in
Fig.~\ref{fig:bcandidates}. The measured distributions of  mass and
transverse momentum of the leading B candidate, \ie that with the
largest $\pt$, as well as $\pt^\cPZ$,  agree with MC predictions within uncertainties.
\begin{figure}[!htbp]
	\begin{center}
\includegraphics[width=0.30\textwidth]{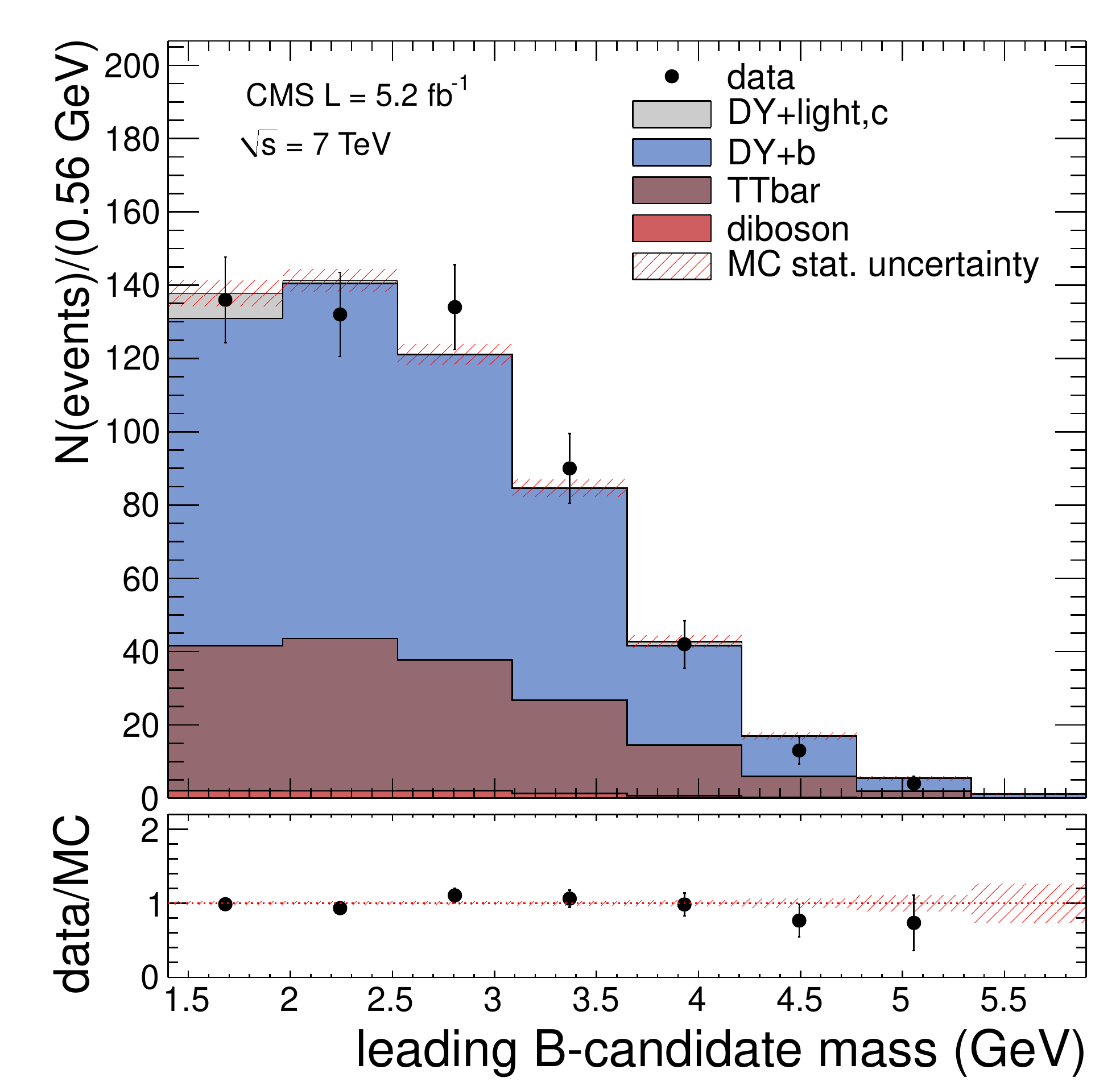}
\includegraphics[width=0.30\textwidth]{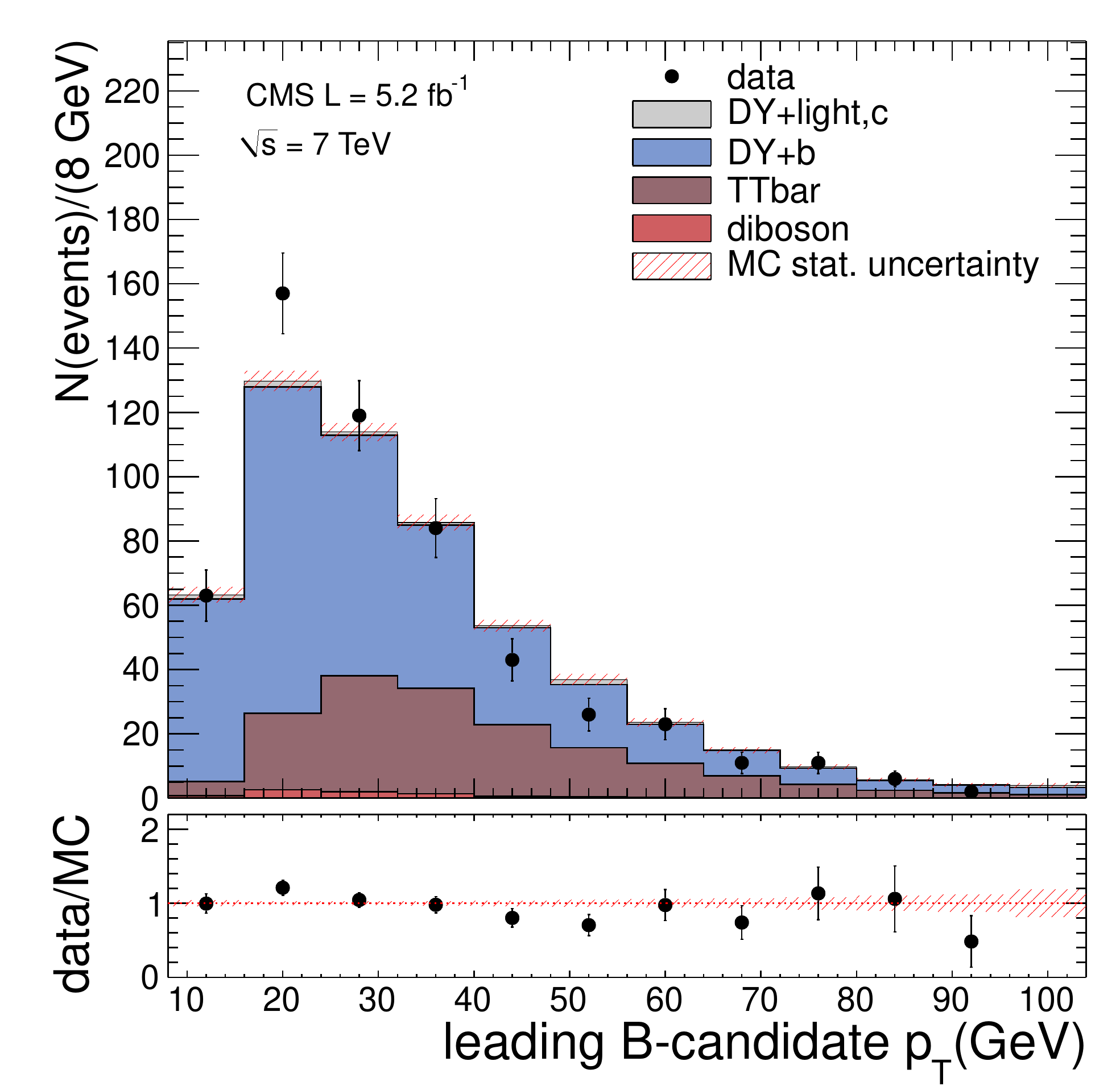}
\includegraphics[width=0.30\textwidth]{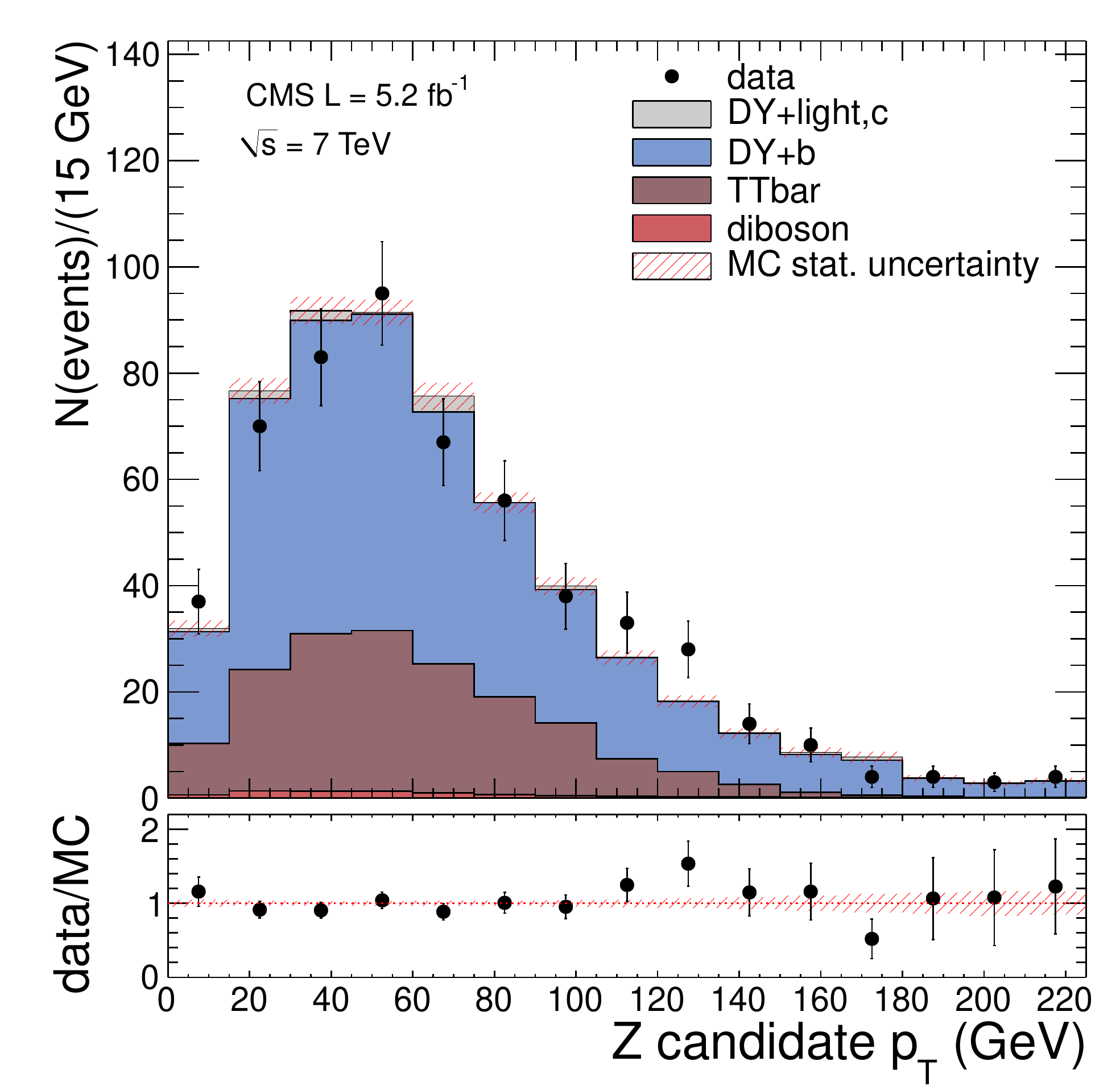}
	\caption{Distribution of the leading B candidate invariant
          mass (left), transverse momentum (centre), and $\pt^\cPZ$
          (right) for the muon and electron channels combined, in the
          signal region ($81<M_{\ell\ell}<101\GeV$). The CMS
          data are represented by solid points and the MC simulation
          by stacked histograms. The shaded region represents the
          statistical uncertainty in the MC prediction. The fraction
          of signal and top background in the simulation is extracted
          by mean a fit (Fig.~\ref{fig:dileptonmassfit}) and the
          sum is normalised to the number of entries in the data. The
          bottom plots show the ratio of measured and simulated numbers
          of entries in each bin with the MC uncertainty represented
          by the dotted area.
	\label{fig:bcandidates}}	
	\end{center}
\end{figure}

\section{Cross section measurement\label{sec:xsec}}

The differential and total cross sections are obtained by subtracting the
background and correcting for
detector acceptance, signal efficiency, and purity. The correction
factors refer to the kinematic phase space for events with exactly two
b hadrons and a lepton pair from a \cPZ~decay. The b hadrons have
$\pt>15\GeV$ and pseudorapidity $\abs{\eta}<2$. Each lepton
has $\pt>20\GeV$, $\abs{\eta}<2.4$, and the dilepton invariant mass is $81<{M}_{\ell
  \ell}<101\GeV$. The differential cross sections are
measured for $\pt^\cPZ>0\GeV$ and $\pt^\cPZ>50\GeV$. In the former
case, the bin sizes are 0.7, 0.53, 0.84, and 0.2 for $\Delta
R_{\mathrm{BB}}$, $\Delta\phi_{\mathrm{BB}}$, min$\Delta
R_{\mathrm{ZB}}$, and $A_{\mathrm{ZBB}}$, respectively. In the latter,
the corresponding values are 0.84, 0.63, 1.0, and 0.25. Since the IVF angular resolution is significantly smaller than the bin size for all
the measured distributions, no unfolding procedure is applied to measure the hadron-level differential cross
sections.

The hadron-level differential cross section is calculated from
\begin{equation}
 \sigma_{\alpha,j} = \mathcal{F}( \mathrm{n}^{\mu}_{\alpha,j}, \mathrm{n}^{\mathrm{e}}_{\alpha,j} )
\cdot \frac{\mathcal{S}^{\mathrm{B}}_{\alpha,j}}{ \epsilon_{\alpha,j}^{2\mathrm{B}}}  \cdot\mathcal{P}_{\alpha,j}
\cdot \frac{1}{\mathcal{L}},
\label{eq:exp-sigma}
\end{equation}
where
\begin{equation}
 \mathrm{n}^{\ell}_{\alpha,j} = \frac{ N^{\ell}_{\alpha,j}}{ \epsilon^{\ell}_{\alpha,j} \cdot A^{\ell}_{\alpha,j} },
 \label{eq:exp-sigmall}
\end{equation}
with $\ell=\Pe,\ \mu$. For each bin~$j$ of the angular variable $\alpha$, indicating one of
the four variables defined in Section~\ref{sec:intro}, the number of signal events
$N_{\alpha,j}^\mathrm{\ell}$ is extracted from an extended unbinned maximum-likelihood fit to the
lepton pair invariant mass distribution. A Breit--Wigner distribution convolved with
a Gaussian resolution function is used for the signal and a third-degree Chebychev polynomial
distribution for the background, as shown in Fig.~\ref{fig:dileptonmassfit}.
The signal shape parameters are evaluated from data while the background parameters are
obtained from simulation. $N_{\alpha,j}^\ell$  is corrected for the dilepton reconstruction and selection
efficiency $\epsilon_{\alpha,j}^{\ell}$ and acceptance  $\mathcal{
  A}_{\alpha,j}^{\ell}$.
The corrected yields $\mathrm{n}^{\ell}_{\alpha,j}$ in the muon and electron channels are found to be in agreement, within statistical uncertainties.

The two channels are combined into a single measurement $ \mathcal{F}(
\mathrm{n}^{\mu}_{\alpha,j}, \mathrm{n}^{\mathrm{e}}_{\alpha,j} )$
using the BLUE algorithm \cite{Lyons:1988rp,Valassi:2003mu}, which performs a weighted average of the input values taking into account the respective uncertainties and their correlations.

The resulting yield is corrected for the b-hadron pair identification efficiency
$\epsilon_{\alpha,j}^{2\mathrm{B}}$, the b-hadron purity $\mathcal{P}_{\alpha,j}$, and the integrated luminosity $\mathcal{L}$. The factor $\mathcal{S}_{\alpha,j}^\mathrm{B}$ corrects for events with b hadrons with $\pt<15\GeV$.

The dilepton trigger efficiency is estimated from data with a
tag-and-probe method, as a
function of the lepton kinematics. It is approximately 93\% for the dimuon and 98\% for
the dielectron trigger selections.
The lepton offline reconstruction and selection efficiencies, around
80\% for muon and 50\% for electron pairs, are obtained from
simulation and are rescaled to match the values measured
in data with a tag-and-probe procedure, as a function of the lepton
pseudorapidity.

The total b-hadron identification efficiency is estimated using multijet events containing semileptonic decays of b-hadrons and
from events enriched with top
quarks. In addition, a dedicated study is
performed to verify that the efficiency measurements are valid for the
inclusive vertex finding algorithm as well.

The efficiency for identifying b-hadron pairs, which ranges
between 8\% and 10\%, is corrected by applying a factor of 0.88 to account for
the discrepancy observed between the measured and simulated
efficiency. This scale factor is measured from data, in the same way
as it is done for the Simple Secondary Vertex method that identifies b
hadrons inside jets~\cite{Chatrchyan:2012jua}. This study requires the
association of the vertices reconstructed with the IVF with jets and
exploits the features of muons produced in
semileptonic decays of the b hadrons, namely their high transverse momenta with respect to the jet axis.
The purity $\mathcal{P}_{\alpha,j}$ and correction factor $S^B_{\alpha,j}$ are evaluated to be about 85\% and 97\%, respectively, based on MC simulation.

The same method is used to derive the total cross section for different ranges of $\pt^\cPZ$. The extended maximum-likelihood fit and the procedure to extract the correction factors are applied to the corresponding event sample.

\section{Systematic uncertainties \label{sec:systematics}}

The following uncertainties on the differential cross sections are considered:
\begin{itemize}
\item{\textit{Uncertainty in combined dilepton signal}}\\
The procedure to combine the muon and electron channels takes into account the systematic
uncertainties on the $N_{\alpha,j}^\ell$ yields and on the dilepton efficiency correction factors.
The systematic uncertainty affecting the resulting combination is estimated by the BLUE
algorithm, and
is approximately ${\pm}2\%$. More details are given below.
\begin{itemize}
\item[--]{\textit{Uncertainty in the signal yield}}\\
The systematic uncertainty associated with the extraction of $N_{\alpha,j}^\ell$ from the extended
unbinned maximum-likelihood fit is estimated by varying the shape parameters within their
uncertainties. For the signal, the shape parameters are the Breit--Wigner mean and width,
as well as the Gaussian standard deviation. For the background, the parameters of the
Chebychev polynomial distribution are considered. A variation of these
factors  leads to a signal yield uncertainty below $\pm$2\%.
\item[--]{\textit{Uncertainty in the trigger efficiency and the lepton efficiency scale
factors}} \\
The lepton reconstruction and selection efficiency corrections are
computed with the MC simulation, and rescaled to match the efficiency values measured from data with the
tag-and-probe method.
The corresponding systematic uncertainty is estimated by varying the scale factors and the
trigger efficiency extracted from data within their systematic uncertainties, mostly due
to the background shape parametrisation.
The resulting variation is $\pm$0.5\% for the muon channel and $\pm$1\% for the electron
channel.
\end{itemize}
\item{\textit{Uncertainty in the efficiency scale factor}}\\
The scale factors between the b-hadron pair identification efficiency in data and simulation
are determined as a function of the jet transverse momentum.
The maximal deviation of the measured values from a constant leads to a $\pm$12\%
systematic uncertainty assigned to the cross section.
\item {\textit{Uncertainty in the purity correction factor}}\\
The purity correction factor accounts for the contamination from events with at least one
reconstructed B candidate produced by a charm hadron decay or, more rarely, by a light
jet. Three categories contribute to such impurity:
$\cPZ\bbbar$ events with a charm hadron from a sequential $\cPqc$ decay reconstructed as b hadron,
$\cPZ\ccbar$ events, and $\cPZ\bbbar\cPqc$ events.
The uncertainty in the purity originates essentially from the
$\cPZ\bbbar\cPqc$ and $\cPZ\ccbar$  processes, where there is
no measurement related to the production of one or two  charm quarks
produced in association with a \cPZ~boson. We therefore provide a
conservative estimate of such uncertainty by varying the
$\cPZ\bbbar\cPqc$ and $\cPZ\ccbar$ fractions by 50\% in the
simulation. The resulting uncertainty in $\mathcal{P}_{\alpha,j}$ is $\pm2.1$\%.
\item{\textit{Bin-to-bin migrations}}\\
Possible migrations of events from one bin to the adjacent ones are
accounted for as a source of systematic uncertainty. The effect varies between
${\pm} 1$--2\% for $\Delta R_{\mathrm{BB}}$ and min$\Delta R_{\mathrm{ZB}}$, and ${\pm} 3$--4\% for the $\Delta
\phi_{\mathrm{BB}}$ and $A_{\mathrm{ZBB}}$ variables.
Such uncertainty does not affect the total cross section measurement.
\item{\textit{Uncertainty in the luminosity}}\\
The luminosity $\mathcal{L}$ is known with a systematic uncertainty of ${\pm}2.2$\%~\cite{CMS:lumi}.
\item {\textit{MC statistical uncertainty}}  \\
The uncertainties on the efficiency and purity corrections are dominated by the limited size of the four-flavour $\cPZ\bbbar$  \MADGRAPH sample. The effect is evaluated in each bin for the differential measurements, and globally for the total cross section determination, and is taken as an additional source
of uncertainty that varies between $\pm2$\% and $\pm3.7$\%.
\end{itemize}

The systematic uncertainties are summarised in Table~\ref{tab:syst}, for the differential
cross sections and  the total cross section measurements.
\begin{table}[htb]
   \centering
  \topcaption{Summary of systematic uncertainties assigned to the
    differential and total
cross section measurements.
  The systematic uncertainties in $N_{\alpha,j}$ and in the
dilepton efficiency are used in the combination of the muon and electron channels, and are
reported in the text. }
\begin{tabular}{lc}
\hline
Source&Uncertainty (\%)\\
\hline
Dilepton channel combination &2 \\
IVF efficiency scale factors & 12 \\
B purity & 2.1 \\
Bin-to-bin migrations ($\Delta R_{\mathrm{BB}}$, min$\Delta
R_{\mathrm{ZB}}$)  & 1--2 \\
Bin-to-bin migrations ($\Delta \phi_{\mathrm{BB}}$, $A_{\mathrm{ZBB}}$) &
3--4 \\
MC statistics --- Differential & 2.0--3.7 \\
MC statistics --- Total & 1.0--3.5  \\
Integrated luminosity & 2.2\\
\hline
\end{tabular}
  \label{tab:syst}               
\end{table}

\section{Theoretical predictions and uncertainties\label{sec:therrors}}

The measured cross sections are compared at hadron level to the predictions
by the \MADGRAPH MC, in both the  five-
(MG5F) and  four-flavour (MG4F)  approaches, and by the \ALPGEN generator in the four-flavour
approach.

The MG5F prediction is based on a matrix-element
calculation where up to four partons are produced in association with
the \cPZ~boson,
the b quarks are assumed massless,
the proton PDF set is  CTEQ6L1, and the jet matching is performed
using the standard \kt-MLM
scheme at a matching scale $\mathcal{
  Q}_\text{match}=20\GeV$~\cite{Alwall:2007fs}. Events with b-hadron pairs
from a second partonic scattering are included.

The MG4F prediction considers  massive b quarks in the matrix-element calculation with the mass set to
 $m_{\cPqb} = 4.7\GeV$. In the matrix element two additional light partons are
 produced in association with the $\cPZ\bbbar$ final state.
The jet matching scheme is also the \kt-MLM  with $\mathcal{
   Q}_\text{match}=30\GeV$.

The \ALPGEN prediction adopts the four-flavour calculation
scheme, with the MLM jet matching and CTEQ5L PDF set. The matching parameters are $\Delta
R^{\text{match}}(\text{parton-jet})=0.7$ and $\pt^{\text{match}}=20\GeV$. In addition to the tree-level predictions mentioned above,
the measurements are compared to the NLO expectations by a\MCATNLO,
which implements the four-flavour scheme with the MSTW2008 NLO PDF set.

The parton shower and hadronisation of all
tree-level samples is obtained with \PYTHIA, with $\pt$-ordered
showers, while a\MCATNLO is interfaced with \HERWIG.
The choices of QCD factorisation and renormalisation scales are summarised in Table~\ref{tabscales}.
\begin{table}
\centering
\topcaption{Summary of the central scale functional forms used in the
  different theoretical predictions for the factorisation ($\mu_F^2$) and
  renormalisation ($\mu_R^2$) scales. The label $jets$ can be (u, d, s, c, b, g) for
  the MG5F production, while it is (u, d, c, s, g) for the MG4F one, for
  which the label b is mentioned explicitly to denote the b
  quark. $m_{\mathrm{T}}$ denotes the transverse mass.\label{tabscales}}
\begin{tabular}{lcc}
\hline
\noalign{\vskip 1mm}
 & $\mu_{F}^2$ & $\mu_{R}^2$\\
\noalign{\vskip 1mm}
\hline
\noalign{\vskip 1mm}
MG5F & $m_\cPZ^2+\pt^2(\text{jets})$ & $\kt^2$ at each
vertex splitting\\
MG4F & $m_{\mathrm{T},\cPZ} \cdot m_\mathrm{T}(\mathrm{b,b})$  & $\kt^2$ at each vertex splitting (excl. b)\\
\ALPGEN & $m_Z^2+
\sum_\text{jets}(m_\text{jets}^2+p_{T,\text{jets}}^2)$ & $\kt^2$ at each vertex splitting (excl. b)\\
a\MCATNLO &
$m_{\ell\ell'}^2+\pt^2(\ell\ell')+\frac{m_{\cPqb}^2+\pt^2(\cPqb)}{2}+\frac{m_{\cPqb}
  '^2+\pt^2(\cPqb')}{2}$ & $=\mu_F^2$\\
\noalign{\vskip 1mm}
\hline
\end{tabular}
\end{table}

The MG5F prediction is rescaled by a $k$-factor of 1.23, corresponding
to the ratio between the next-to-next-to-leading-order (NNLO) prediction of the inclusive \cPZ~production
cross section, and the tree-level cross section from
\MADGRAPH. The tree-level cross section prediction for MG4F (\ALPGEN)
is rescaled by a k-factor obtained from the a\MCATNLO cross section
of 16\unit{pb} obtained for $M_{\ell\ell}>30\GeV$ divided by the
corresponding MG4F (\ALPGEN) prediction.

The following uncertainties on the theoretical predictions are
considered and combined quadratically:
\begin{itemize}
\item The shape uncertainties associated with the b-quark mass,
  $m_\cPqb$, for the
  \MADGRAPH 4F prediction are assessed by varying  $m_\cPqb$
  between 4.4 and 5.0\GeV. Each distribution is rescaled so that the normalisation matches the NLO cross section provided by a\MCATNLO and the envelope is considered as the uncertainty band.
\item The shape uncertainties due to the factorisation and renormalisation
scales are assessed for the  \MADGRAPH
4F and 5F predictions by varying their values simultaneously by a factor of two. The \MADGRAPH
4F (5F) distributions are rescaled so that the normalisation matches the NLO (NNLO) cross section provided by a\MCATNLO (FEWZ~\cite{Gavin:2010az}) and the envelopes are considered as uncertainty bands.
\item The uncertainties associated with the matching scale are assessed by
  varying it by ${\pm}15\%$ for \MADGRAPH 4F and by a factor of
  two for the 5F case.
\item The shape uncertainties associated with the choice of PDF set are found to
  be negligible. The effect of PDF variations are included as
  normalisation uncertainties as described in the next item.
\item For \MADGRAPH 4F and \ALPGEN predictions the normalisation uncertainty is given by
  the corresponding a\MCATNLO cross section uncertainty. The latter is
  obtained by varying the factorisation and renormalisation scales
  simultaneously by a factor of two, and by replacing the MSTW2008 PDF
  set with CT10. For \MADGRAPH
  5F the normalisation uncertainty is given by the corresponding NNLO
  cross section uncertainty~\cite{Gavin:2010az}.
\item For a\MCATNLO the uncertainty associated with the parton shower is
  assessed from the difference between \PYTHIA (D6T tune) with
  virtuality-ordered showers and \HERWIG.
\item The statistical uncertainty due to the finite size of the
  simulated sample is propagated for all theoretical predictions.
\end{itemize}

\section{Results \label{sec:results}}

The measured differential cross sections as a function of  the three angular variables and the angular asymmetry variable are shown in Figs.~\ref{fig:diffxsec}
and~\ref{fig:diffxsec50} for all $\pt^\cPZ$ and for $\pt^\cPZ>50\GeV$, respectively.
Figure~\ref{fig:diffxsec} shows that the  $\Delta R_{\mathrm{BB}}$
collinear region is better described by \ALPGEN, while the four- and
five-flavour \MADGRAPH as well as a\MCATNLO predictions tend to
underestimate the data.  
At large $\Delta R_{\mathrm{BB}}$, all predictions are
in good agreement with the data. The fraction of the cross section
with collinear b hadrons increases for $\pt^\cPZ>50\GeV$ and in
this case \ALPGEN also gives the best description of the measured distributions.

Similar conclusions can be drawn from the $\Delta\phi_{\mathrm{BB}}$
distribution. In the nonboosted case, data are above all MC
predictions in the region of back-to-back b-hadron pairs by
approximately one standard deviation.
This discrepancy vanishes for $\pt^\cPZ>50\GeV$.
The simulated min$\Delta R_{\mathrm{ZB}}$ and $A_{\mathrm{ZBB}}$
generally agree with the data. Some discrepancy is observed at min$\Delta
R_{\mathrm{ZB}}>2$ in both ranges of $\pt^\cPZ$, and at low
$A_\mathrm{ZBB}$.
The data are found to be above the predictions primarily in the regions
where the contributions from the $\cPq {i} \rightarrow \cPZ
\bbbar \mathrm{X}$ subprocesses are expected to be dominant, as shown
in Fig.~\ref{plotth}.
\begin{figure}[h!tb]
\begin{center}
\includegraphics[width=0.48\textwidth]{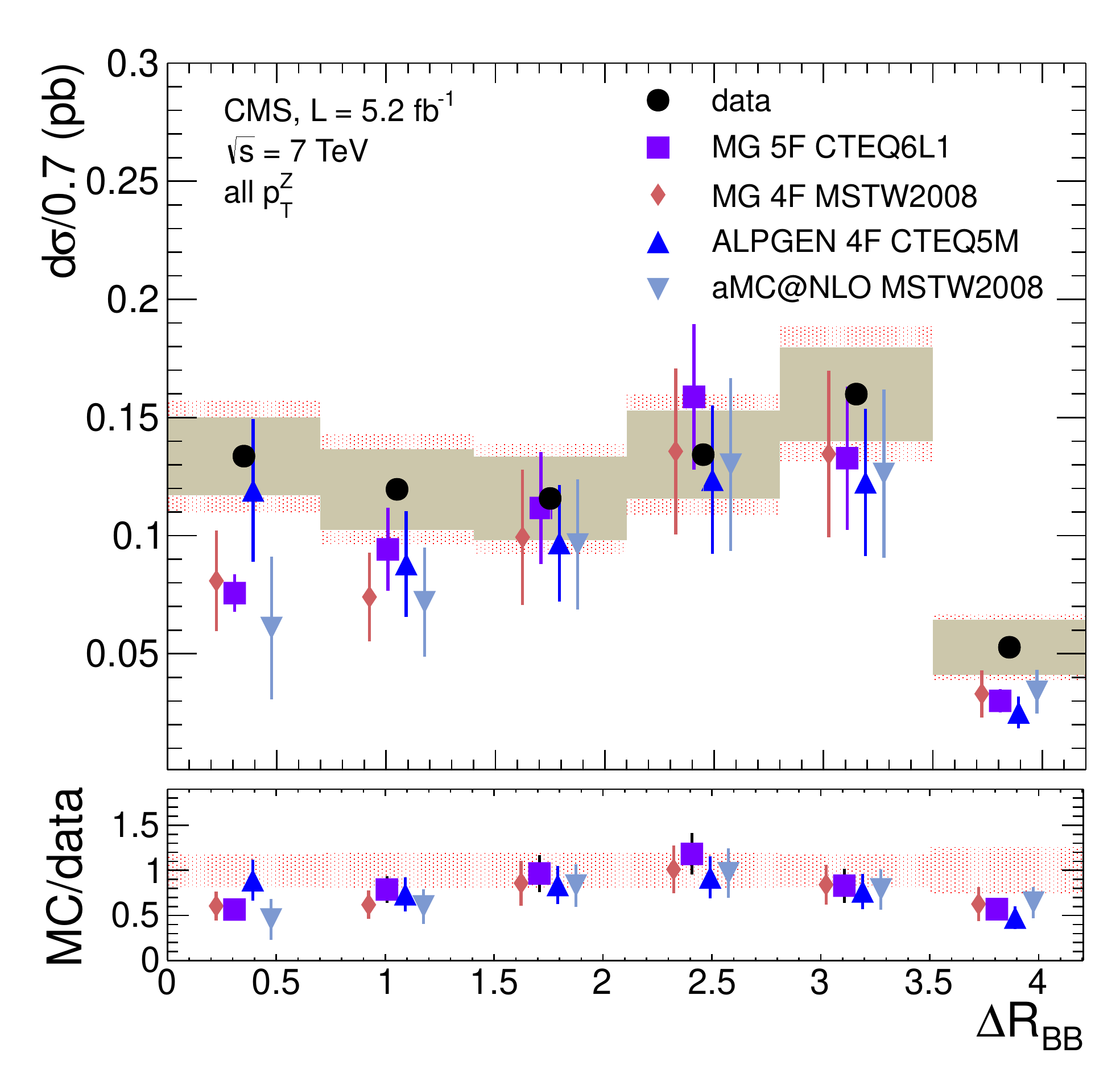}
\includegraphics[width=0.48\textwidth]{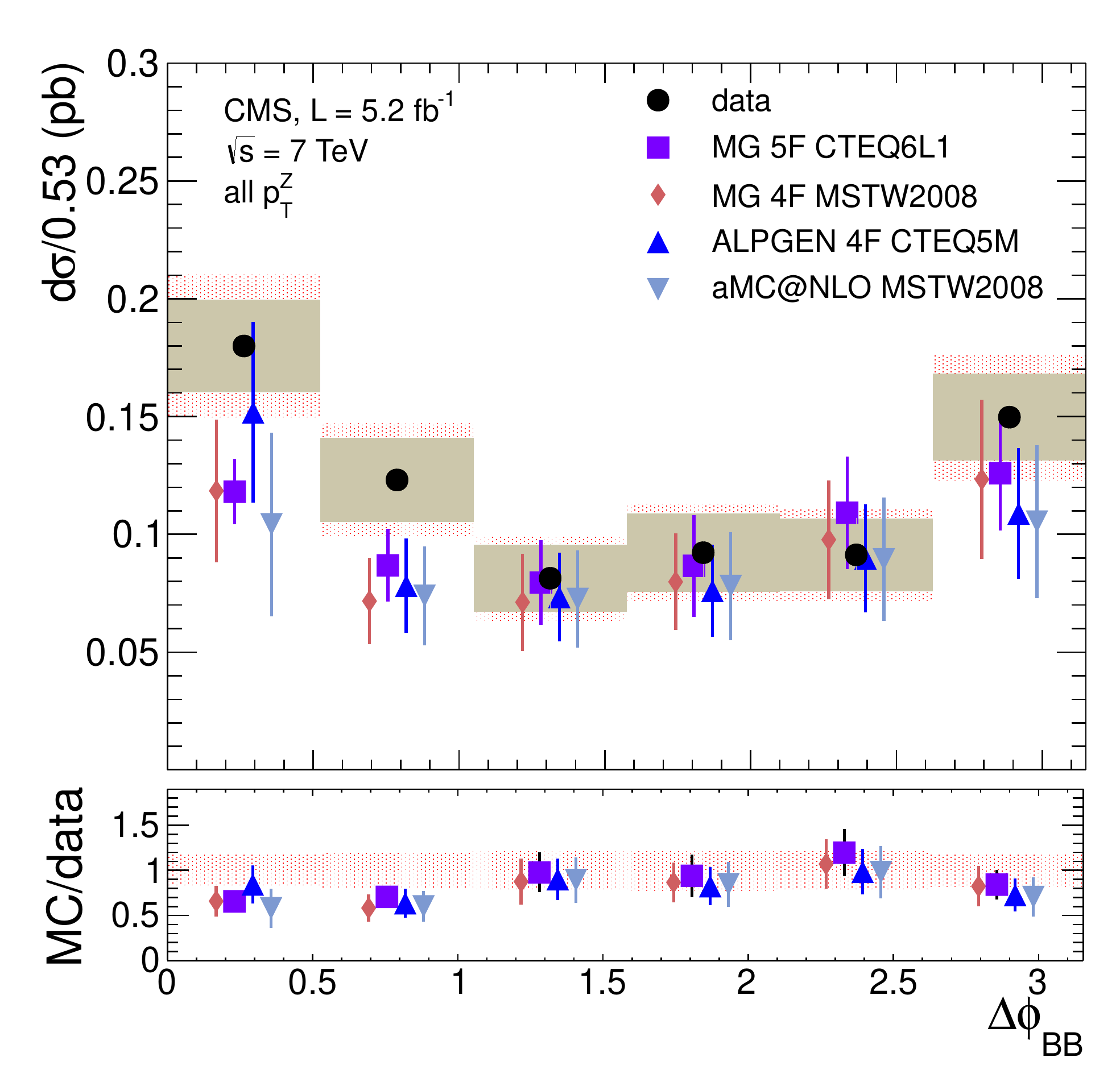}
\includegraphics[width=0.48\textwidth]{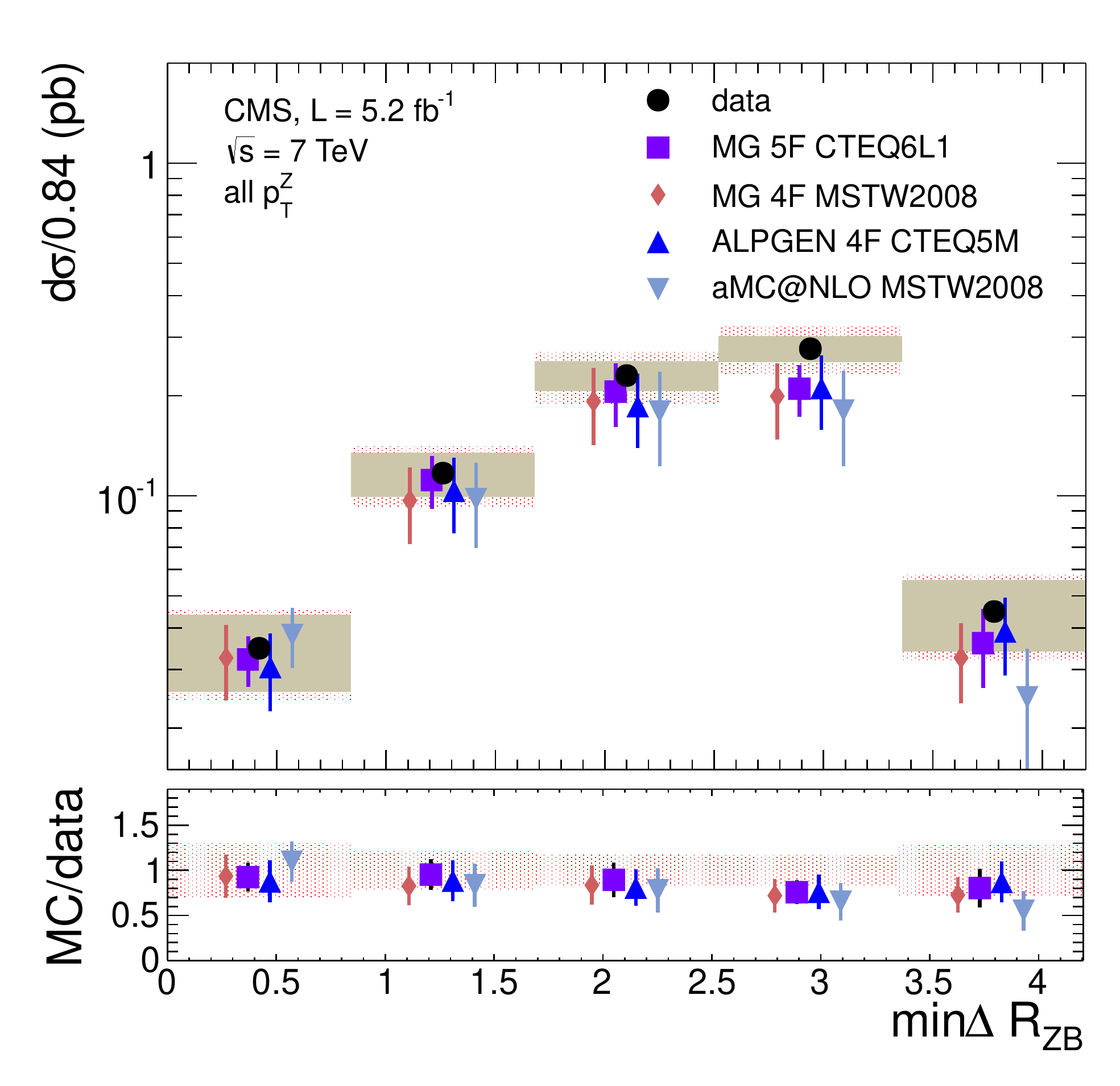}
\includegraphics[width=0.48\textwidth]{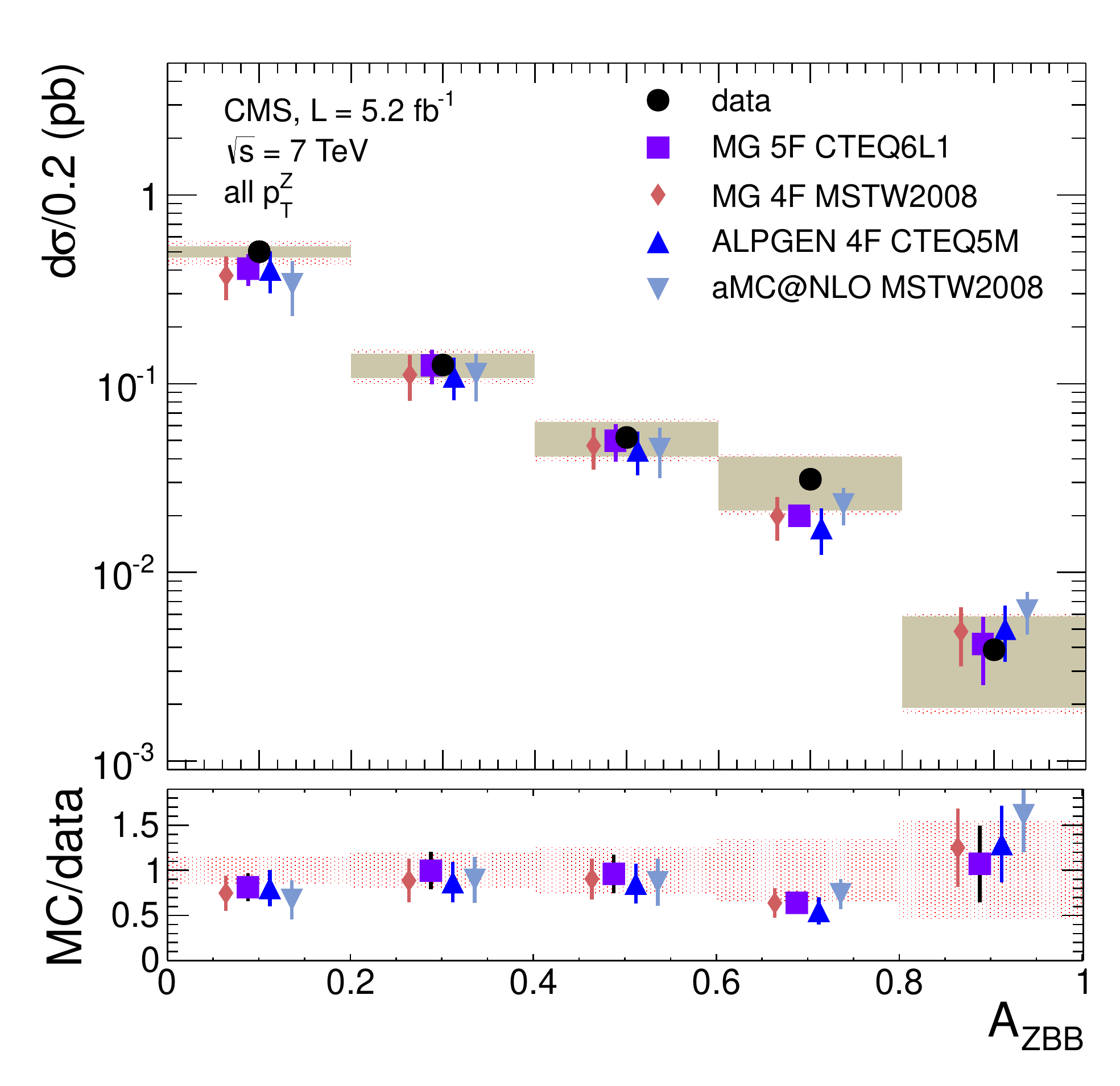}
\caption{Differential cross sections for all $\pt^\cPZ$, as a
function of $\Delta R_{\mathrm{BB}}$ (top left), $\Delta \phi_{\mathrm{BB}}$
(top right), min$\Delta R_{\mathrm{ZB}}$ (bottom left),
and $A_{\mathrm{ZBB}}$ (bottom right). The measured values are shown as
black points. The dotted bands correspond to the quadratic sum of statistical and
systematic uncertainties. Statistical uncertainties are shown
separately as solid bands. The measurements are compared to the hadron-level predictions by \MADGRAPH
in the four- and
five-flavour schemes, \ALPGEN, and a\MCATNLO. For each distribution  the ratio between the Monte Carlo
predictions and the measurements is also shown, with the total experimental
uncertainty indicated by the dotted area.
\label{fig:diffxsec}}
\end{center}
\end{figure}
\begin{figure}[h!tb]
\begin{center}
\includegraphics[width=0.48\textwidth]{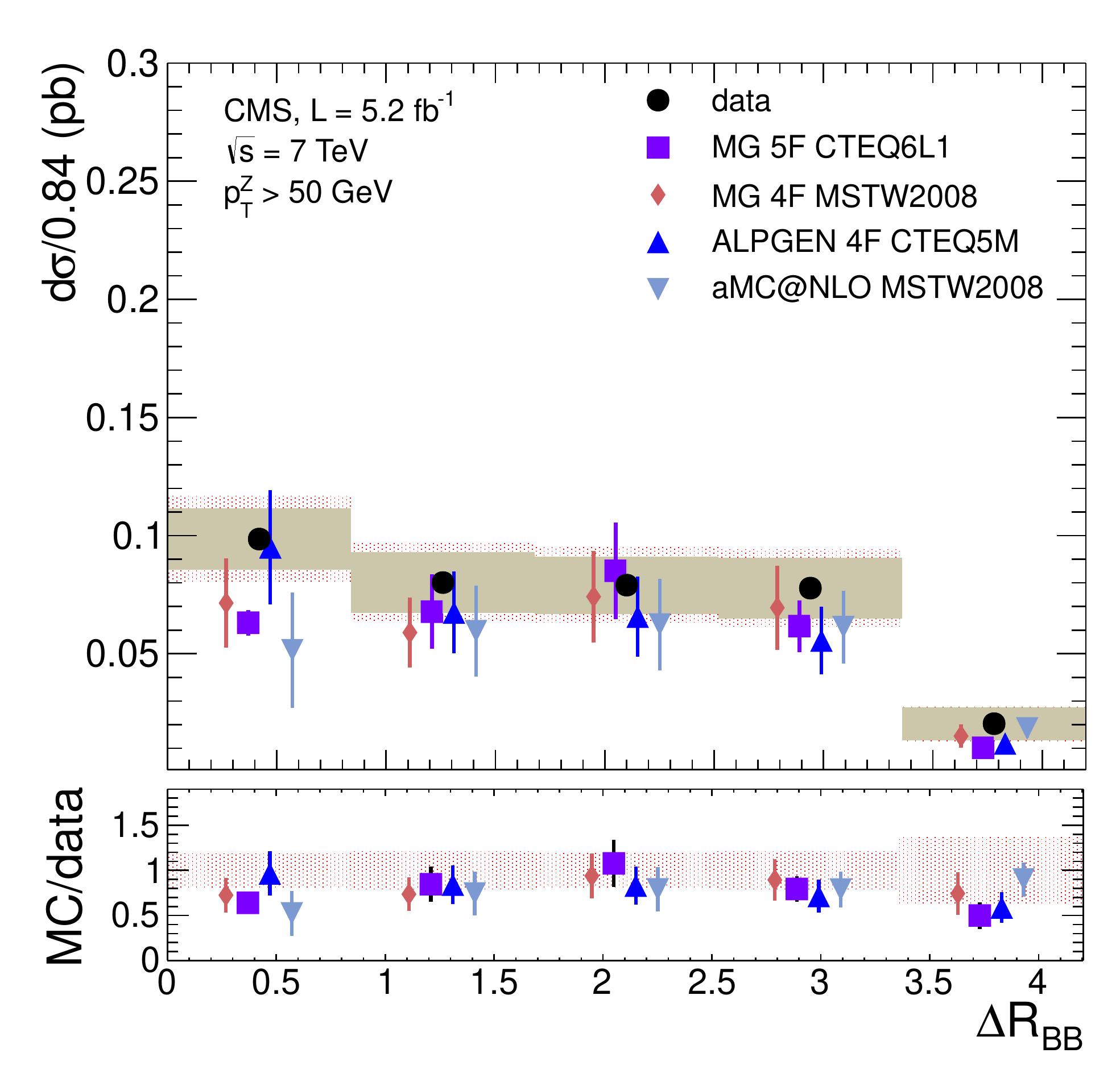}
\includegraphics[width=0.48\textwidth]{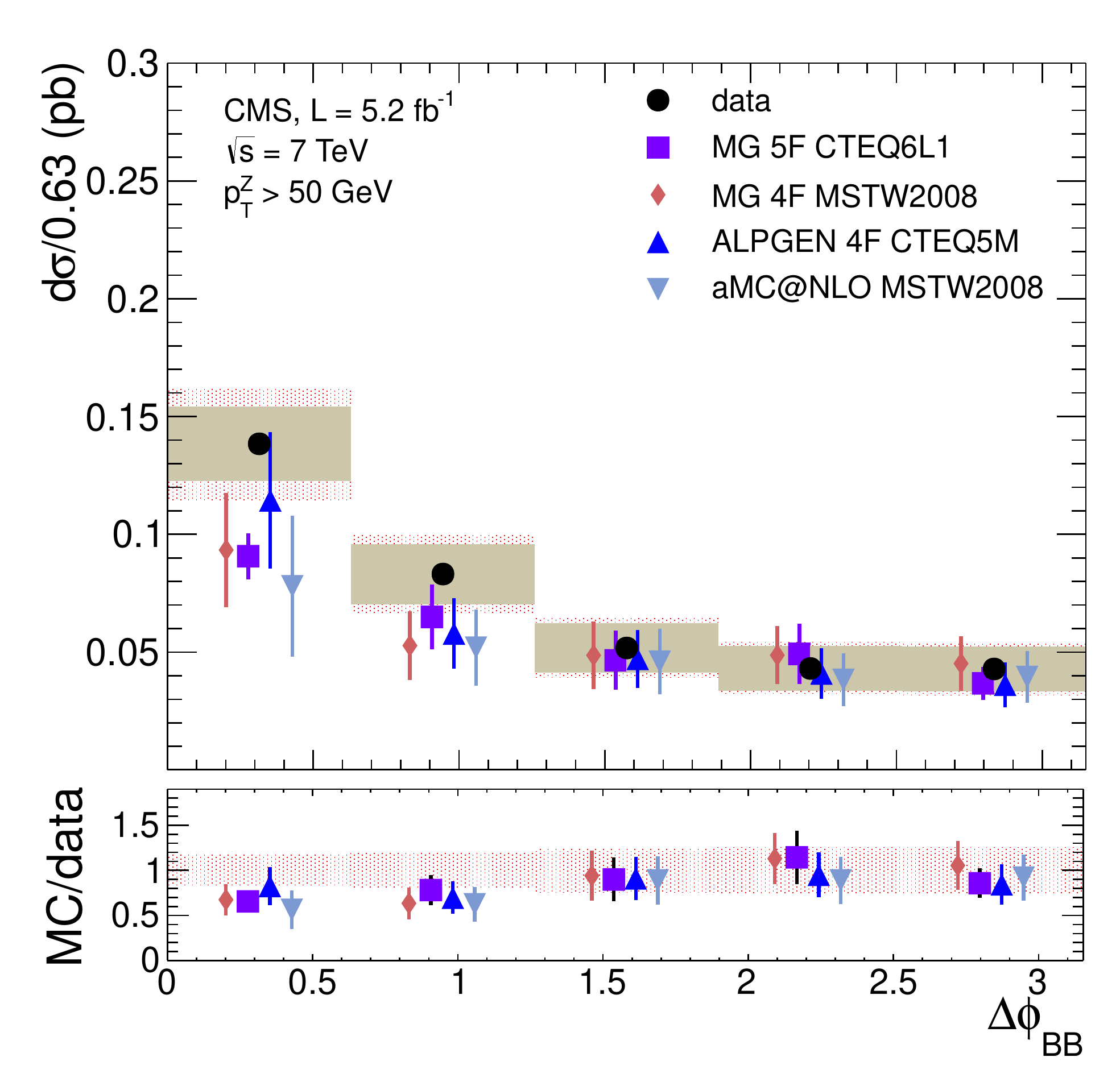}
\includegraphics[width=0.48\textwidth]{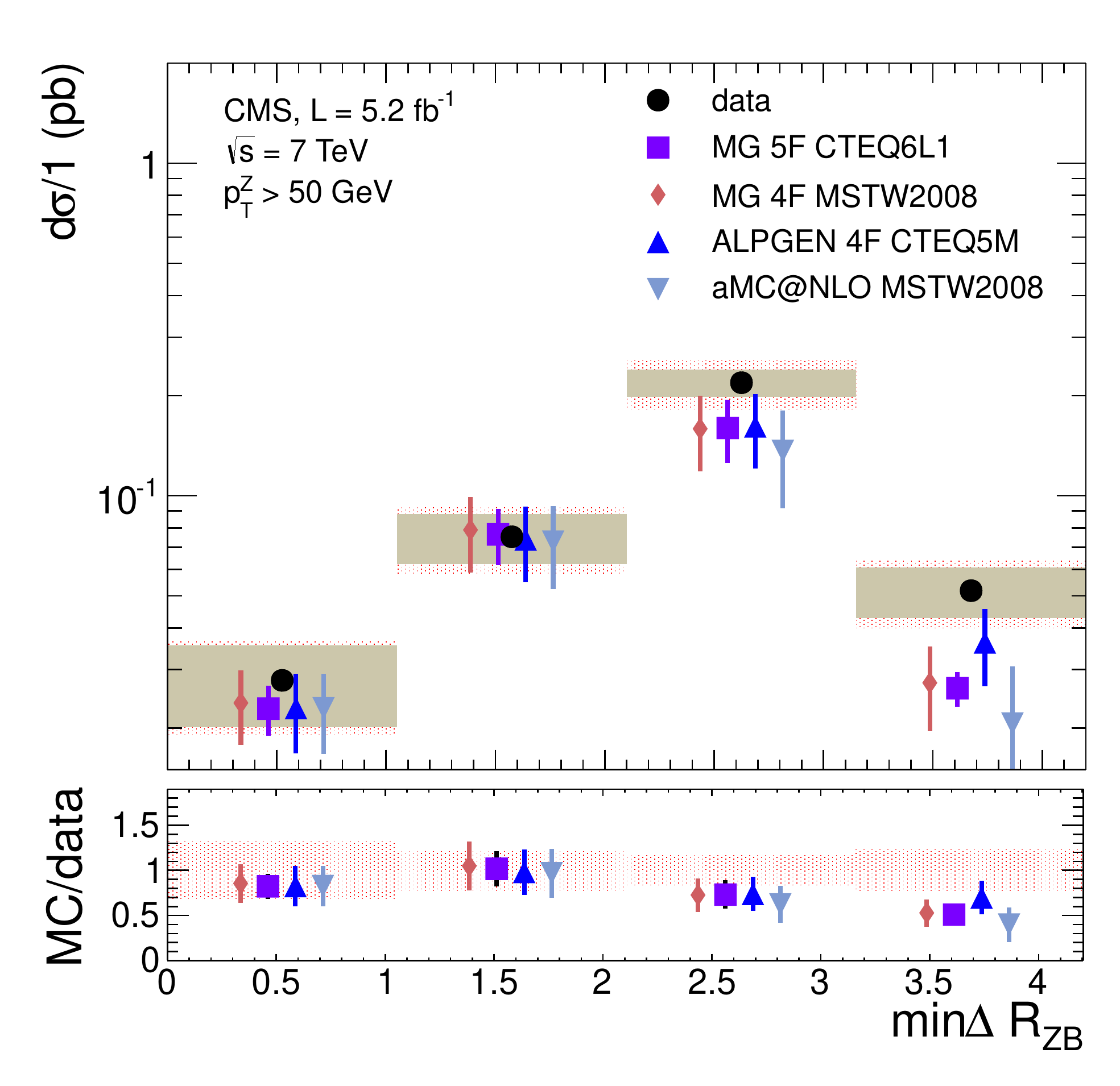}
\includegraphics[width=0.48\textwidth]{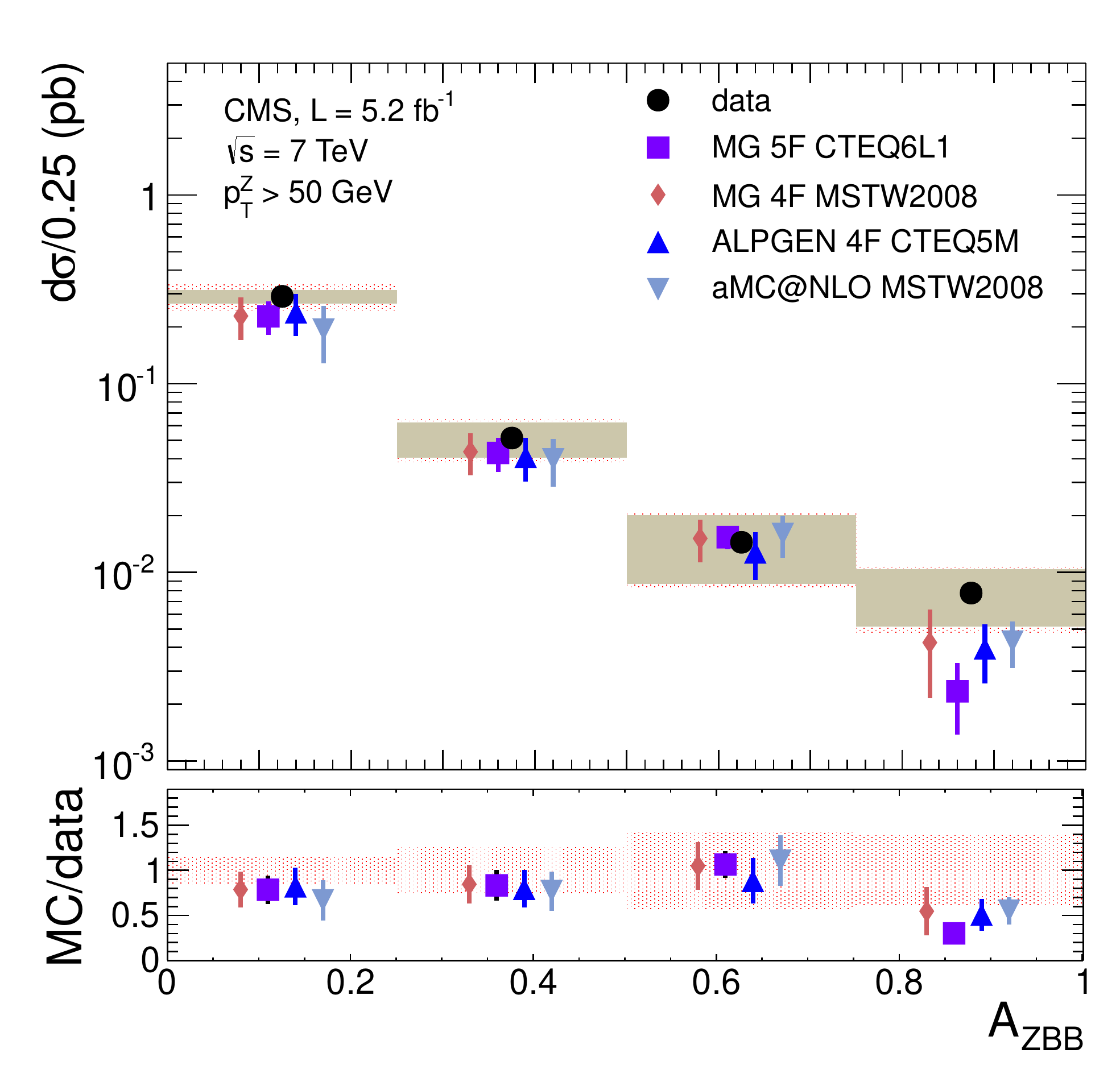}
\caption{Differential cross sections for $\pt^\cPZ>50\GeV$, as a
function of $\Delta R_{\mathrm{BB}}$ (top left), $\Delta \phi_{\mathrm{BB}}$
(top right), min$\Delta R_{\mathrm{ZB}}$ (bottom left),
and $A_{\mathrm{ZBB}}$ (bottom right). The measured values are shown as
black points. The dotted bands correspond to the quadratic sum of statistical and
systematic uncertainties. Statistical uncertainties are shown
separately as solid bands. The measurements are compared to the hadron-level predictions by \MADGRAPH
in the four- and
five-flavour schemes, \ALPGEN, and a\MCATNLO. For each distribution  the ratio between the Monte Carlo
predictions and the measurements is also shown, with the total experimental
uncertainty indicated by the dotted area.
\label{fig:diffxsec50}}
\end{center}
\end{figure}

The total hadron-level cross section is shown in
Fig.~\ref{fig:xsecvsptZ} for four different regions of  $\pt^\cPZ$: for the inclusive
spectrum, and for $\pt^\cPZ>40$, 80, and 120\GeV.  Data points are
generally above all simulations by  about 15\%, apart from
a\MCATNLO for which the discrepancy can be as large as 50\%  at large $\pt^\cPZ$.
\begin{figure}[h!tb]
\begin{center}
\includegraphics[width=0.50\textwidth]{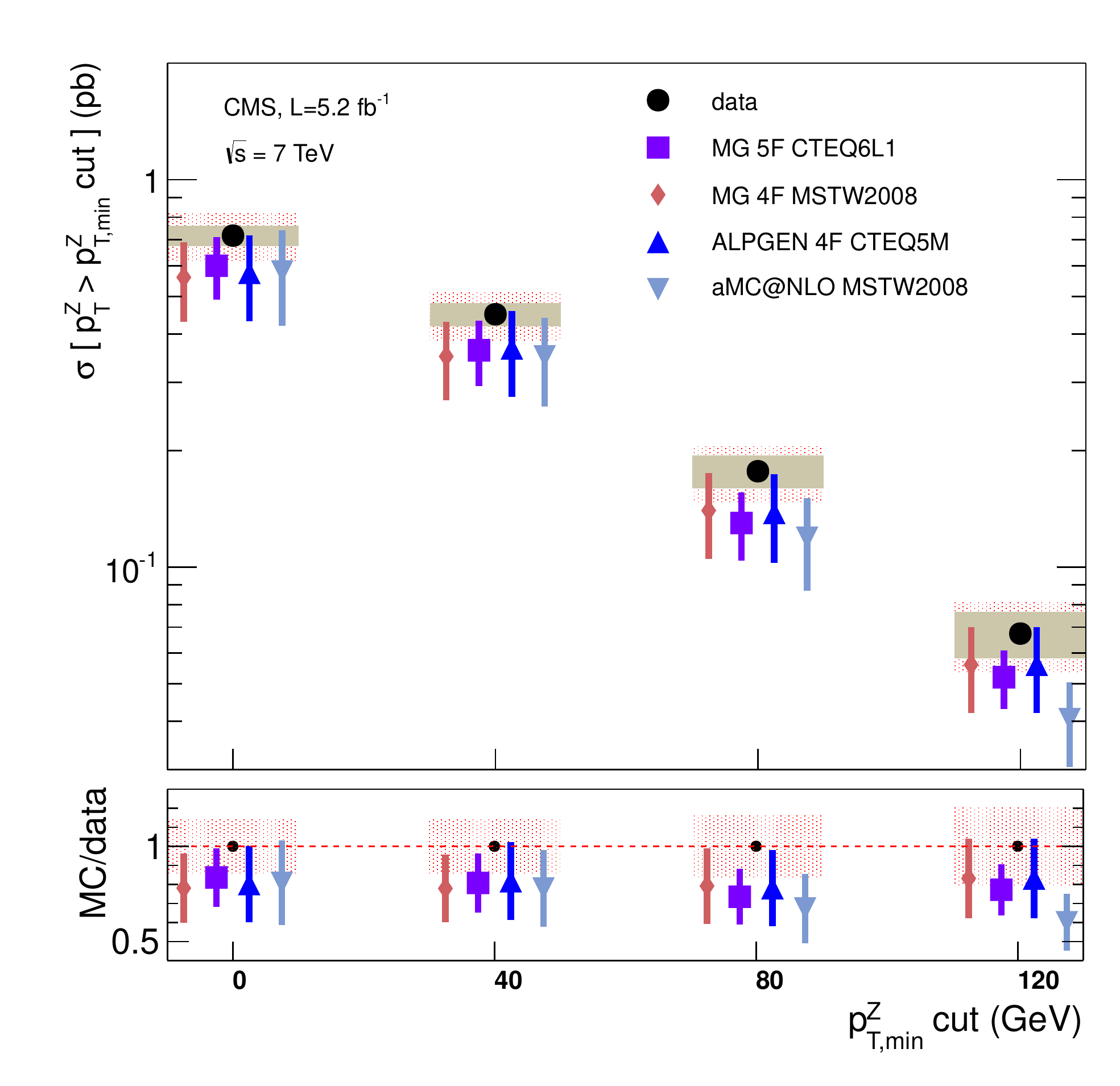}
\caption{
Total cross section as function of the cut on $\pt^\cPZ$. The
measured values are
shown as black points. The solid bands include statistical and systematic
uncertainties combined quadratically. Statistical uncertainties are
shown separately as dotted bands. The measurements are compared to the hadron-level predictions by \MADGRAPH
in the four- and five-flavour scheme, \ALPGEN, and a\MCATNLO. The ratio between the Monte Carlo
predictions and the measurements is also shown, with the total experimental
uncertainty indicated by the dotted area.
\label{fig:xsecvsptZ}}
\end{center}
\end{figure}
\section{Conclusions \label{sec:conclusions}}

The first measurement of angular correlations in the process $\mathrm{pp}\rightarrow\cPZ\bbbar X$ has been performed. The analysed data set corresponds to an integrated
luminosity of 5.1\fbinv recorded by the CMS experiment in 2011 at a
centre-of-mass energy of 7\TeV. \cPZ~bosons are reconstructed in the $\Pep\Pem$ and $\mu^+\mu^-$ decay modes. The use of the
inclusive vertex finder, which exploits the excellent CMS tracking
performance, allows the full angular range to be probed, including
configurations with collinear b hadrons.

The production cross sections are measured as functions of four
angular variables:
$\Delta R_{\mathrm{BB}}$, $\Delta\phi_{\mathrm{BB}}$, min$\Delta
R_{\mathrm{ZB}}$, and $A_{\mathrm{ZBB}}$.
The measurements are compared with tree-level predictions by the \MADGRAPH and
\ALPGEN MC generators implementing different flavour number schemes.
The variables most sensitive to the b-hadron production process,
$\Delta R_\mathrm{BB}$ and $\Delta\phi_\mathrm{BB}$, show that the four-flavour
prediction implemented in \ALPGEN provides the best description of CMS data.

The MG5F MC generator has been one of the standard tools used to
simulate backgrounds from associated production of vector bosons and
heavy quarks for Higgs boson and new physics searches as well as
SM studies. The results reported here indicate that such a
description may not be optimal for analyses sensitive to
the production of collinear b hadrons. This fact may be
particularly important in the simulation of the $\PW\bbbar$ process,
where collinear b-hadron production is expected to be enhanced compared to the $\cPZ\bbbar$
process.

This is the first time that a\MCATNLO predictions, in which QCD
contributions are computed to NLO,
have been compared with data for the $\cPZ\bbbar$ process. It is found that a\MCATNLO
underestimates the cross section at low $\Delta R_\mathrm{BB}$ and
$\Delta\phi_\mathrm{BB}$, and at large  min$\Delta R_{\mathrm{ZB}}$.
A comprehensive assessment of the
a\MCATNLO predictions requires further studies of the
scale choices and parton shower modelling.
It is worth noting that the use of NLO jet matching would also
improve the precision of the prediction at small values of $\Delta
R_\mathrm{BB}$.

The total hadron-level cross section $\sigma_{\text{tot}} =
\sigma(\Pp\Pp\to\cPZ\bbbar X)\mathcal{B}(\cPZ \rightarrow \ell^+ \ell^-)$ is also evaluated in different
ranges of the \cPZ~boson transverse momentum.
For the case where no
cut is applied on the \cPZ~momentum, the total cross section is
$\sigma_{\text{tot}}=0.71\pm0.08$\unit{pb};
for $\pt^\cPZ>40\GeV$,  $\sigma_{\text{tot}}=0.44\pm0.05$\unit{pb};
for $\pt^\cPZ>80\GeV$,  $\sigma_{\text{tot}}=0.17\pm0.02$\unit{pb};
and for  $\pt^\cPZ>120\GeV$,  $\sigma_{\text{tot}}=0.07\pm0.01$\unit{pb}.
The measured values are systematically larger than MC predictions,
partly because of the excess observed in the collinear $\Delta
R_\mathrm{BB}$ region. The shape of the measured
integrated cross section as a function of the minimum transverse
momentum of the \cPZ~boson is in good agreement with the tree-level 4F
predictions, while slightly larger discrepancies are observed for MG5F and even more for a\MCATNLO, particularly at large $\pt^\cPZ$.

\section*{Acknowledgements}
We thank Marco Zaro for his dedication in simulating  a\MCATNLO
samples for this analysis.

\hyphenation{Bundes-ministerium Forschungs-gemeinschaft Forschungs-zentren} We congratulate our colleagues in the CERN accelerator departments for the excellent performance of the LHC and thank the technical and administrative staffs at CERN and at other CMS institutes for their contributions to the success of the CMS effort. In addition, we gratefully acknowledge the computing centres and personnel of the Worldwide LHC Computing Grid for delivering so effectively the computing infrastructure essential to our analyses. Finally, we acknowledge the enduring support for the construction and operation of the LHC and the CMS detector provided by the following funding agencies: the Austrian Federal Ministry of Science and Research and the Austrian Science Fund; the Belgian Fonds de la Recherche Scientifique, and Fonds voor Wetenschappelijk Onderzoek; the Brazilian Funding Agencies (CNPq, CAPES, FAPERJ, and FAPESP); the Bulgarian Ministry of Education, Youth and Science; CERN; the Chinese Academy of Sciences, Ministry of Science and Technology, and National Natural Science Foundation of China; the Colombian Funding Agency (COLCIENCIAS); the Croatian Ministry of Science, Education and Sport; the Research Promotion Foundation, Cyprus; the Ministry of Education and Research, Recurrent financing contract SF0690030s09 and European Regional Development Fund, Estonia; the Academy of Finland, Finnish Ministry of Education and Culture, and Helsinki Institute of Physics; the Institut National de Physique Nucl\'eaire et de Physique des Particules~/~CNRS, and Commissariat \`a l'\'Energie Atomique et aux \'Energies Alternatives~/~CEA, France; the Bundesministerium f\"ur Bildung und Forschung, Deutsche Forschungsgemeinschaft, and Helmholtz-Gemeinschaft Deutscher Forschungszentren, Germany; the General Secretariat for Research and Technology, Greece; the National Scientific Research Foundation, and National Office for Research and Technology, Hungary; the Department of Atomic Energy and the Department of Science and Technology, India; the Institute for Studies in Theoretical Physics and Mathematics, Iran; the Science Foundation, Ireland; the Istituto Nazionale di Fisica Nucleare, Italy; the Korean Ministry of Education, Science and Technology and the World Class University program of NRF, Republic of Korea; the Lithuanian Academy of Sciences; the Mexican Funding Agencies (CINVESTAV, CONACYT, SEP, and UASLP-FAI); the Ministry of Science and Innovation, New Zealand; the Pakistan Atomic Energy Commission; the Ministry of Science and Higher Education and the National Science Centre, Poland; the Funda\c{c}\~ao para a Ci\^encia e a Tecnologia, Portugal; JINR (Armenia, Belarus, Georgia, Ukraine, Uzbekistan); the Ministry of Education and Science of the Russian Federation, the Federal Agency of Atomic Energy of the Russian Federation, Russian Academy of Sciences, and the Russian Foundation for Basic Research; the Ministry of Science and Technological Development of Serbia; the Secretar\'{\i}a de Estado de Investigaci\'on, Desarrollo e Innovaci\'on and Programa Consolider-Ingenio 2010, Spain; the Swiss Funding Agencies (ETH Board, ETH Zurich, PSI, SNF, UniZH, Canton Zurich, and SER); the National Science Council, Taipei; the Thailand Center of Excellence in Physics, the Institute for the Promotion of Teaching Science and Technology of Thailand and the National Science and Technology Development Agency of Thailand; the Scientific and Technical Research Council of Turkey, and Turkish Atomic Energy Authority; the Science and Technology Facilities Council, UK; the US Department of Energy, and the US National Science Foundation.

Individuals have received support from the Marie-Curie programme and the European Research Council and EPLANET (European Union); the Leventis Foundation; the A. P. Sloan Foundation; the Alexander von Humboldt Foundation; the Belgian Federal Science Policy Office; the Fonds pour la Formation \`a la Recherche dans l'Industrie et dans l'Agriculture (FRIA-Belgium); the Agentschap voor Innovatie door Wetenschap en Technologie (IWT-Belgium); the Ministry of Education, Youth and Sports (MEYS) of Czech Republic; the Council of Science and Industrial Research, India; the Compagnia di San Paolo (Torino); the HOMING PLUS programme of Foundation for Polish Science, cofinanced by EU, Regional Development Fund; and the Thalis and Aristeia programmes cofinanced by EU-ESF and the Greek NSRF.

\bibliography{auto_generated}   

\providecommand{\href}[2]{#2}\begingroup\raggedright\begin{thebibliography}{10}%
\makeatletter
\providecommand{\hrefCMSnoop }[0]{\@secondoftwo}%
\makeatother
\providecommand{\doi}{\texttt{doi:}\begingroup \urlstyle{tt}\Url}

\bibitem{Aad20121}
\hrefCMSnoop {} {{ ATLAS} Collaboration, ``Observation of a new particle in the
  search for the Standard Model Higgs boson with the ATLAS detector at the
  LHC'',} \textit{ Phys. Lett. B} \textbf{ 716} (2012) 1,
  \href{http://dx.doi.org/10.1016/j.physletb.2012.08.020}{\doi{10.1016/j.physletb.2012.08.020}}.

\bibitem{:2012gu}
\hrefCMSnoop {} {{ CMS} Collaboration, ``{Observation of a new boson at a mass
  of 125 GeV with the CMS experiment at the LHC}'',} \textit{ Phys. Lett. B}
  \textbf{ 716} (2012) 30,
  \href{http://dx.doi.org/10.1016/j.physletb.2012.08.021}{\doi{10.1016/j.physletb.2012.08.021}},
\href{http://www.arXiv.org/abs/1207.7235}{\texttt{ arXiv:1207.7235}}.

\bibitem{Gerard:2007kn}
\hrefCMSnoop {} {J.-M. G{\'e}rard and M.~Herquet, ``{Twisted Custodial Symmetry
  in Two-Higgs-Doublet Models}'',} \textit{ Phys. Rev. Lett.} \textbf{ 98}
  (2007) 251802,
  \href{http://dx.doi.org/10.1103/PhysRevLett.98.251802}{\doi{10.1103/PhysRevLett.98.251802}},
\href{http://www.arXiv.org/abs/hep-ph/0703051}{\texttt{ arXiv:hep-ph/0703051}}.

\bibitem{deVisscher:2009zb}
S.~de~Visscher\hrefCMSnoop {} { {et~al.}, ``{Unconventional phenomenology of a
  minimal two-Higgs-doublet model}'',} \textit{ JHEP} \textbf{ 08} (2009) 042,
  \href{http://dx.doi.org/10.1088/1126-6708/2009/08/042}{\doi{10.1088/1126-6708/2009/08/042}},
\href{http://www.arXiv.org/abs/0904.0705}{\texttt{ arXiv:0904.0705}}.

\bibitem{Dermisek:2008id}
\hrefCMSnoop {} {R.~Dermisek, ``{Light CP-odd Higgs and Small tan beta Scenario
  in the MSSM and Beyond}'',} (2008).
\href{http://www.arXiv.org/abs/0806.0847}{\texttt{ arXiv:0806.0847}}.

\bibitem{Dermisek:2008uu}
\hrefCMSnoop {} {R.~Dermisek and J.~F. Gunion, ``{Many light Higgs bosons in
  the next-to-minimal supersymmetric model}'',} \textit{ Phys. Rev. D} \textbf{
  79} (2009) 055014,
  \href{http://dx.doi.org/10.1103/PhysRevD.79.055014}{\doi{10.1103/PhysRevD.79.055014}},
\href{http://www.arXiv.org/abs/0811.3537}{\texttt{ arXiv:0811.3537}}.

\bibitem{Alwall:2007st}
J.~Alwall\hrefCMSnoop {} { {et~al.}, ``{MadGraph/MadEvent v4: the new web
  generation}'',} \textit{ JHEP} \textbf{ 09} (2007) 028,
  \href{http://dx.doi.org/10.1088/1126-6708/2007/09/028}{\doi{10.1088/1126-6708/2007/09/028}},
\href{http://www.arXiv.org/abs/0706.2334}{\texttt{ arXiv:0706.2334}}.

\bibitem{Alwall:2011uj}
J.~Alwall\hrefCMSnoop {} { {et~al.}, ``{MadGraph 5: going beyond}'',} \textit{
  JHEP} \textbf{ 06} (2011) 128,
  \href{http://dx.doi.org/10.1007/JHEP06(2011)128}{\doi{10.1007/JHEP06(2011)128}},
\href{http://www.arXiv.org/abs/1106.0522}{\texttt{ arXiv:1106.0522}}.

\bibitem{alpgen}
M.~L. Mangano\hrefCMSnoop {} { {et~al.}, ``{ALPGEN, a generator for hard
  multiparton processes in hadronic collisions}'',} \textit{ JHEP} \textbf{ 07}
  (2003) 001,
  \href{http://dx.doi.org/10.1088/1126-6708/2003/07/001}{\doi{10.1088/1126-6708/2003/07/001}},
\href{http://www.arXiv.org/abs/hep-ph/0206293}{\texttt{ arXiv:hep-ph/0206293}}.

\bibitem{Gleisberg:2008ta}
T.~Gleisberg\hrefCMSnoop {} { {et~al.}, ``{Event generation with SHERPA
  1.1}'',} \textit{ JHEP} \textbf{ 02} (2009) 007,
  \href{http://dx.doi.org/10.1088/1126-6708/2009/02/007}{\doi{10.1088/1126-6708/2009/02/007}},
\href{http://www.arXiv.org/abs/0811.4622}{\texttt{ arXiv:0811.4622}}.

\bibitem{Campbell:2000bg}
\hrefCMSnoop {} {J.~M. Campbell and R.~K. Ellis, ``{Radiative corrections to
  $\cPZ\bbbar$ production}'',} \textit{ Phys. Rev. D} \textbf{ 62} (2000)
  114012,
  \href{http://dx.doi.org/10.1103/PhysRevD.62.114012}{\doi{10.1103/PhysRevD.62.114012}},
\href{http://www.arXiv.org/abs/hep-ph/0006304}{\texttt{ arXiv:hep-ph/0006304}}.

\bibitem{FebresCordero:2008ci}
\hrefCMSnoop {} {F.~Febres~Cordero, L.~Reina, and D.~Wackeroth, ``{NLO QCD
  corrections to $\cPZ\bbbar$ production with massive bottom quarks at the
  Fermilab Tevatron}'',} \textit{ Phys. Rev. D} \textbf{ 78} (2008) 074014,
  \href{http://dx.doi.org/10.1103/PhysRevD.78.074014}{\doi{10.1103/PhysRevD.78.074014}},
\href{http://www.arXiv.org/abs/0806.0808}{\texttt{ arXiv:0806.0808}}.

\bibitem{Cordero:2009kv}
\hrefCMSnoop {} {F.~Febres~Cordero, L.~Reina, and D.~Wackeroth, ``{W- and
  Z-boson production with a massive bottom-quark pair at the Large Hadron
  Collider}'',} \textit{ Phys. Rev. D} \textbf{ 80} (2009) 034015,
  \href{http://dx.doi.org/10.1103/PhysRevD.80.034015}{\doi{10.1103/PhysRevD.80.034015}},
\href{http://www.arXiv.org/abs/0906.1923}{\texttt{ arXiv:0906.1923}}.

\bibitem{Frederix:2011qg}
R.~Frederix\hrefCMSnoop {} { {et~al.}, ``{W and $\cPZ/\gamma*$ boson production
  in association with a bottom-antibottom pair}'',} \textit{ JHEP} \textbf{ 09}
  (2011) 061,
  \href{http://dx.doi.org/10.1007/JHEP09(2011)061}{\doi{10.1007/JHEP09(2011)061}},
\href{http://www.arXiv.org/abs/1106.6019}{\texttt{ arXiv:1106.6019}}.

\bibitem{Frixione:2002ik}
\hrefCMSnoop {} {S.~Frixione and B.~R. Webber, ``{Matching NLO QCD computations
  and parton shower simulations}'',} \textit{ JHEP} \textbf{ 06} (2002) 029,
  \href{http://dx.doi.org/10.1088/1126-6708/2002/06/029}{\doi{10.1088/1126-6708/2002/06/029}},
\href{http://www.arXiv.org/abs/hep-ph/0204244}{\texttt{ arXiv:hep-ph/0204244}}.

\bibitem{Maltoni:2012pa}
\hrefCMSnoop {} {F.~Maltoni, G.~Ridolfi, and M.~Ubiali, ``{b-initiated
  processes at the LHC: a reappraisal}'',} \textit{ JHEP} \textbf{ 07} (2012)
  022,
  \href{http://dx.doi.org/10.1007/JHEP04(2013)095}{\doi{10.1007/JHEP04(2013)095}},
\href{http://www.arXiv.org/abs/1203.6393}{\texttt{ arXiv:1203.6393}}.

\bibitem{Chatrchyan:2012jua}
\hrefCMSnoop {} {{ CMS} Collaboration, ``{Identification of b-quark jets with
  the CMS experiment}'',} \textit{ JINST} \textbf{ 8} (2013) P04013,
  \href{http://dx.doi.org/10.1088/1748-0221/8/04/P04013}{\doi{10.1088/1748-0221/8/04/P04013}},
\href{http://www.arXiv.org/abs/1211.4462}{\texttt{ arXiv:1211.4462}}.

\bibitem{Khachatryan:2011wq}
\hrefCMSnoop {} {{ CMS} Collaboration, ``{Measurement of
  $\mathrm{B\overline{B}}$ angular correlations based on secondary vertex
  reconstruction at $\sqrt{s}=7\TeV$}'',} \textit{ JHEP} \textbf{ 03} (2011)
  136,
  \href{http://dx.doi.org/10.1007/JHEP03(2011)136}{\doi{10.1007/JHEP03(2011)136}},
\href{http://www.arXiv.org/abs/1102.3194}{\texttt{ arXiv:1102.3194}}.

\bibitem{Rubin:2010xp}
\hrefCMSnoop {} {M.~Rubin, G.~P. Salam, and S.~Sapeta, ``{Giant QCD K-factors
  beyond NLO}'',} \textit{ JHEP} \textbf{ 09} (2010) 084,
  \href{http://dx.doi.org/10.1007/JHEP09(2010)084}{\doi{10.1007/JHEP09(2010)084}},
\href{http://www.arXiv.org/abs/1006.2144}{\texttt{ arXiv:1006.2144}}.

\bibitem{Aad:2011jn}
\hrefCMSnoop {} {{ ATLAS} Collaboration, ``{Measurement of the cross-section
  for $b-$jets produced in association with a \cPZ\ boson at $\sqrt{s}=7\TeV$
  with the ATLAS detector}'',} \textit{ Phys. Lett. B} \textbf{ 706} (2012)
  295,
  \href{http://dx.doi.org/10.1016/j.physletb.2011.11.059}{\doi{10.1016/j.physletb.2011.11.059}},
\href{http://www.arXiv.org/abs/1109.1403}{\texttt{ arXiv:1109.1403}}.

\bibitem{Chatrchyan:2012vr}
\hrefCMSnoop {} {{ CMS} Collaboration, ``{Measurement of the
  $\cPZ/\gamma^*+$b-jet cross section in pp collisions at $\sqrt{s}=7\TeV$}'',}
  \textit{ JHEP} \textbf{ 06} (2012) 126,
  \href{http://dx.doi.org/10.1007/JHEP06(2012)126}{\doi{10.1007/JHEP06(2012)126}},
\href{http://www.arXiv.org/abs/1204.1643}{\texttt{ arXiv:1204.1643}}.

\bibitem{Chatrchyan:2008aa}
\hrefCMSnoop {} {{ CMS} Collaboration, ``{The CMS experiment at the CERN
  LHC}'',} \textit{ JINST} \textbf{ 3} (2008) S08004,
\href{http://dx.doi.org/10.1088/1748-0221/3/08/S08004}{\doi{10.1088/1748-0221/3/08/S08004}}.

\bibitem{Agostinelli:2002hh}
\hrefCMSnoop {} {{ GEANT4} Collaboration, ``{\GEANT 4--a simulation
  toolkit}'',} \textit{ Nucl. Instrum. Meth. A} \textbf{ 506} (2003) 250,
\href{http://dx.doi.org/10.1016/S0168-9002(03)01368-8}{\doi{10.1016/S0168-9002(03)01368-8}}.

\bibitem{Sjostrand:2006za}
\hrefCMSnoop {} {T.~Sj{\"o}strand, S.~Mrenna, and P.~Z. Skands, ``{\PYTHIA 6.4
  physics and manual}'',} \textit{ JHEP} \textbf{ 05} (2006) 026,
  \href{http://dx.doi.org/10.1088/1126-6708/2006/05/026}{\doi{10.1088/1126-6708/2006/05/026}},
\href{http://www.arXiv.org/abs/hep-ph/0603175}{\texttt{ arXiv:hep-ph/0603175}}.

\bibitem{Field:2010bc}
\hrefCMSnoop {} {R.~Field, ``{Early LHC Underlying Event Data - Findings and
  Surprises}'',} (2010).
\href{http://www.arXiv.org/abs/1010.3558}{\texttt{ arXiv:1010.3558}}.

\bibitem{Chatrchyan:2012xi}
\hrefCMSnoop {} {{ CMS} Collaboration, ``{Performance of CMS muon
  reconstruction in pp collision events at $\sqrt{s}=7\TeV$}'',} \textit{
  JINST} \textbf{ 7} (2012) P10002,
  \href{http://dx.doi.org/10.1088/1748-0221/7/10/P10002}{\doi{10.1088/1748-0221/7/10/P10002}},
\href{http://www.arXiv.org/abs/1206.4071}{\texttt{ arXiv:1206.4071}}.

\bibitem{Khachatryan:2010xn}
\hrefCMSnoop {} {{ CMS} Collaboration, ``{Measurements of inclusive W~and
  \cPZ~cross sections in pp collisions at $\sqrt{s}=7\TeV$}'',} \textit{ JHEP}
  \textbf{ 01} (2011) 080,
  \href{http://dx.doi.org/10.1007/JHEP01(2011)080}{\doi{10.1007/JHEP01(2011)080}},
\href{http://www.arXiv.org/abs/1012.2466}{\texttt{ arXiv:1012.2466}}.

\bibitem{CMS:2009nxa}
\href {http://cdsweb.cern.ch/record/1194487} {{ CMS} Collaboration,
  ``{Particle-Flow Event Reconstruction in CMS and Performance for Jets, Taus,
  and $E_T^\mathrm{miss}$}'',} CMS Physics Analysis Summary CMS-PAS-PFT-09-001,
  (2009).

\bibitem{Cacciari:2007fd}
\hrefCMSnoop {} {M.~Cacciari and G.~P. Salam, ``{Pileup subtraction using jet
  areas}'',} \textit{ Phys. Lett. B} \textbf{ 659} (2008) 119,
  \href{http://dx.doi.org/10.1016/j.physletb.2007.09.077}{\doi{10.1016/j.physletb.2007.09.077}},
\href{http://www.arXiv.org/abs/0707.1378}{\texttt{ arXiv:0707.1378}}.

\bibitem{Fruhwirth:2007hz}
\hrefCMSnoop {} {W.~Waltenberger, R.~Fr{\"u}hwirth, , and P.~Vanlaer,
  ``{Adaptive vertex fitting}'',} \textit{ J. Phys. G} \textbf{ 34} (2007)
  N343,
\href{http://dx.doi.org/10.1088/0954-3899/34/12/N01}{\doi{10.1088/0954-3899/34/12/N01}}.

\bibitem{Lyons:1988rp}
\hrefCMSnoop {} {L.~Lyons, D.~Gibaut, and P.~Clifford, ``{How to combine
  correlated estimates of a single physical quantity}'',} \textit{ Nucl.
  Instrum. Meth. A} \textbf{ 270} (1988) 110,
\href{http://dx.doi.org/10.1016/0168-9002(88)90018-6}{\doi{10.1016/0168-9002(88)90018-6}}.

\bibitem{Valassi:2003mu}
\hrefCMSnoop {} {A.~Valassi, ``{Combining correlated measurements of several
  different physical quantities}'',} \textit{ Nucl. Instrum. Meth. A} \textbf{
  500} (2003) 391,
\href{http://dx.doi.org/10.1016/S0168-9002(03)00329-2}{\doi{10.1016/S0168-9002(03)00329-2}}.

\bibitem{CMS:lumi}
\href {http://cdsweb.cern.ch/record/1434360} {{CMS Collaboration}, ``Absolute
  Calibration of the Luminosity Measurement at CMS: Winter 2012 Update'',} CMS
  Physics Analysis Summary CMS-PAS-SMP-12-008, (2012).

\bibitem{Alwall:2007fs}
J.~Alwall\hrefCMSnoop {} { {et~al.}, ``{Comparative study of various algorithms
  for the merging of parton showers and matrix elements in hadronic
  collisions}'',} \textit{ Eur. Phys. J. C} \textbf{ 53} (2008) 473,
  \href{http://dx.doi.org/10.1140/epjc/s10052-007-0490-5}{\doi{10.1140/epjc/s10052-007-0490-5}},
\href{http://www.arXiv.org/abs/0706.2569}{\texttt{ arXiv:0706.2569}}.

\bibitem{Gavin:2010az}
\hrefCMSnoop {} {R.~Gavin, Y.~Li, F.~Petriello, and S.~Quackenbush, ``{FEWZ
  2.0: A code for hadronic Z production at next-to-next-to-leading order}'',}
  \textit{ Comput. Phys. Commun.} \textbf{ 182} (2011) 2388,
  \href{http://dx.doi.org/10.1016/j.cpc.2011.06.008}{\doi{10.1016/j.cpc.2011.06.008}},
\href{http://www.arXiv.org/abs/1011.3540}{\texttt{ arXiv:1011.3540}}.

\end{thebibliography}\endgroup

\cleardoublepage \appendix\section{The CMS Collaboration \label{app:collab}}\begin{sloppypar}\hyphenpenalty=5000\widowpenalty=500\clubpenalty=5000\textbf{Yerevan Physics Institute,  Yerevan,  Armenia}\\*[0pt]
S.~Chatrchyan, V.~Khachatryan, A.M.~Sirunyan, A.~Tumasyan
\vskip\cmsinstskip
\textbf{Institut f\"{u}r Hochenergiephysik der OeAW,  Wien,  Austria}\\*[0pt]
W.~Adam, T.~Bergauer, M.~Dragicevic, J.~Er\"{o}, C.~Fabjan\cmsAuthorMark{1}, M.~Friedl, R.~Fr\"{u}hwirth\cmsAuthorMark{1}, V.M.~Ghete, N.~H\"{o}rmann, J.~Hrubec, M.~Jeitler\cmsAuthorMark{1}, W.~Kiesenhofer, V.~Kn\"{u}nz, M.~Krammer\cmsAuthorMark{1}, I.~Kr\"{a}tschmer, D.~Liko, I.~Mikulec, D.~Rabady\cmsAuthorMark{2}, B.~Rahbaran, C.~Rohringer, H.~Rohringer, R.~Sch\"{o}fbeck, J.~Strauss, A.~Taurok, W.~Treberer-Treberspurg, W.~Waltenberger, C.-E.~Wulz\cmsAuthorMark{1}
% ...
\vskip\cmsinstskip
% ...
1:~~Also at Vienna University of Technology, Vienna, Austria\\
2:~~Also at CERN, European Organization for Nuclear Research, Geneva, Switzerland\\
% ...

\end{sloppypar}
\end{document}